\documentclass[preprint,aps,12pt,showpacs,nofootinbib,tightenlines]{revtex4}
 \usepackage{amsmath}
 \usepackage{amssymb}
 \usepackage{epsfig}
 \usepackage{graphicx}
\begin{document}

\newcommand{\beq}{\begin{eqnarray}}
\newcommand{\eeq}{\end{eqnarray}}
\def\bea{\begin{eqnarray}}
\def\eea{\end{eqnarray}}
\newcommand{\bvga}{ B\to V \gamma }
\newcommand{\bsga}{b\to s \gamma}
\newcommand{\bsg}{b\to s g}
\newcommand{\bdga}{b\to d \gamma}
\newcommand{\bksga}{B\to K^* \gamma}
\newcommand{\brhoga}{B\to \rho \gamma}

\newcommand{\bxsga}{B\to X_s \gamma}
\newcommand{\brbsga}{{\cal B}(b\to s \gamma)}
\newcommand{\brbxsga}{{\cal B}(B\to X_s \gamma)}
\newcommand{\bzbzb}{ B_d^0 - \bar{B}_d^0 }
\newcommand{\mt}{m_t}
\newcommand{\mw}{M_W}
\newcommand{\mh}{M_H}
\newcommand{\mz}{M_Z}
\newcommand{\muw}{\mu_W}
\newcommand{\mub}{\mu_b}

\newcommand{\mhp}{M_{H^\pm}}
\newcommand{\mha}{M_{A}}
\newcommand{\mhz}{M_{h^0}}
\newcommand{\dmd}{\Delta M_{B_d} }
\newcommand{\dms}{\Delta M_{B_s} }
\newcommand{\hbbd}{\hat{B}_{B_d} }
\newcommand{\fbbds}{ f^2_{B_d} \hat{B}_{B_d} }
\newcommand{\fbbd}{ f_{B_d} \sqrt{\hat{B}_{B_d} } }
\newcommand{\fbbs}{ f_{B_s} \sqrt{\hat{B}_{B_s} } }

\newcommand{\ltt}{\lambda_{tt} }
\newcommand{\lbb}{\lambda_{bb} }
\newcommand{\rhob}{\bar{\rho} }
\newcommand{\etab}{\bar{\eta} }

\newcommand{\tab}[1]{Table \ref{#1}}
\newcommand{\fig}[1]{Fig.\ref{#1}}
\newcommand{\real}{{\rm Re}\,}
\newcommand{\im}{{\rm Im}\,}
\newcommand{\non}{\nonumber\\ }
\newcommand{\calb}{{\cal B} }
\newcommand{\as}{\alpha_{\scriptscriptstyle S}}
\newcommand{\dbars}{  |\overline{D}|^2}

\newcommand{\smallsm}{{\scriptscriptstyle SM}}
\newcommand{\smallyy}{{\scriptscriptstyle YY}}
\newcommand{\smallxy}{{\scriptscriptstyle XY}}
\newcommand{\smallnp}{{\scriptscriptstyle NP}}

\def \epjc{  Eur. Phys. J. C }
\def \jpg{  J. Phys. G }
\def \jhep{  J. High Energy Phys. }

\def \npb{  Nucl. Phys. B }
\def \plb{  Phys. Lett. B }
\def \prd{  Phys. Rev. D }
\def \prl{  Phys. Rev. Lett.  }
\def \pr{   Phys. Rep. }
\def \rmp{  Rev. Mod. Phys. }
%
\title{$B^0-\bar{B}^0$ mixing and  $B \to X_s \gamma $ decay in the third type 2HDM:
effects of NLO QCD contributions}
\author{ Zhenjun Xiao}
\email{zjxiao@email.njnu.edu.cn;    xiaozj@ictp.trieste.it}
\affiliation{Department of Physics, Nanjing Normal University,
Nanjing, Jiangsu 210097, P.R.China}
\affiliation{High Energy Section, ICTP, Strada Costiera 11, 34014 Trieste,
Italy}

\author{Libo Guo}
\email{guolb@email.njnu.edu.cn}
\affiliation{Department of Physics, Nanjing Normal University,
Nanjing, Jiangsu 210097, P.R.China}
\date{\today}
\begin{abstract}
In this paper, we calculated the next-to-leading order (NLO) new physics contributions
to the mass splitting $\dmd$
and the branching ratio $\brbxsga$ induced by the charged Higgs loop diagrams
in the third type of two-Higgs-doublet models (model III) and draw the constraints on the
free parameters of model III. For the model III under consideration, we found that
(a) an upper limit $|\ltt|\leq 1.7$ is obtained from the precision data of
$\dmd=0.502 \pm 0.007 ps^{-1}$, while $|\ltt| \approx 0.5$ is favored phenomenologicaly;
(b) for $B \to X_s \gamma$ decay, the NLO QCD contributions tend to cancel the
LO new physics contributions;
(c) a light charged Higgs boson with a mass around or even less than
200 GeV is still allowed at NLO level by the measured branching ratio $\brbxsga$:
numerically, $188 \leq \mh \leq 215$ GeV for $(|\ltt|,|\lbb|)=(0.5,18)$;
(d) the NLO QCD contributions tend to cancel the LO contributions effectively,
the lower limit on $\mh$ is consequently decreased by about $200$ GeV;
(e) the allowed region of $\mh$ will be shifted toward heavy mass
end for a non-zero relative phase $\theta$ between the Yukawa couplings $\ltt$ and $\lbb$.
The numerical results for the conventional model II
are also presented  for the sake of a comparison.
\end{abstract}

\pacs{13.20.He, 12.60.Fr, 14.40.Nd}

\maketitle

\newpage

\section{introduction}\label{sec-1}

Among physical observables of B meson mixing and decays, the mass splitting
$\dmd$ and the branching ratio $\brbxsga$ have been measured with high precision.
Recent world averages as given in Refs.\cite{jessop02,pdg2003,hfag03} are
\beq
\calb (\bxsga) &=& (3.34 \pm 0.38) \times 10^{-4}, \label{eq:bsga-exp}\\
\dmd &=& 0.502 \pm 0.007\, ps^{-1}, \label{eq:dmd-exp}
\eeq
which are in perfect agreement with the next-to-leading order standard model (SM)
predictions, for example, presented in Refs.\cite{bg98,urban98}.
Obviously, there is only a small room left for new physics effects beyond the SM.
By comparing the theoretical predictions with the precision data, strong
constraints on the parameter space of new physics models can be obtained.

During the past decade, the $B \to X_s \gamma$ decay has been studied in great detail
in the SM and various new physics models.
At present, the complete NLO calculations of this decay mode are available for the SM
\cite{cmm97,kagan99,buras02,gm01,hurth02,greub03}, for the conventional two-Higgs doublet models
(2HDM) \cite{bg98,ciuchini98,crs98,gm01}, for some supersymmetric models \cite{susy,bobeth},
and for left-right symmetric models \cite{bobeth}. The studies at leading order (LO)  for
model III \cite{atwood96,atwood97,chao99,xiao00}, for the top-two-Higgs-doublet
model \cite{wu99}  and the Technicolor models
\cite{xiao96} are also available.
From relevant theoretical calculations, we know that the rare decay $B \to X_s \gamma$ is
very sensitive to the new physics contributions and has been used as the first test for
new physics models.

In Ref.\cite{bg98}, Borzumati and Greub studied the $B \to X_s \gamma$ decay in
NLO QCD in a class of models containing at least two Higgs doublets and with
only one charged Higgs boson nondecoupled at low energy,  and found the
constraints on the mass of charged-Higgs boson for model II: the popular type-II
two-Higgs-doublet model. In this paper, we will extend  their work \cite{bg98}
to the case of model III ( the third type two-Higgs-doublet model\cite{hou92}), to constrain
the charged Higgs boson mass $\mh$, as well as the the Yuakawa couplings  $\ltt$ and
$\lbb$ after  the inclusion of the NLO QCD corrections.

The strength of the $B_d^0-\overline{B}_d^0$ mixing is described by the
mass splitting $\dmd$. In the SM, $\dmd$ is strongly dominated by the box diagrams
involving the heavy top quark and W gauge boson. In new physics models,
the box diagrams where one or two $W$ gauge bosons are replaced by other charged particles
can also contribute to this quantity. The well measured $\dmd$ therefore can be
used to constrain the new physics models.

In Ref.\cite{urban98}, the authors presented a calculation of the mass
splitting $\dmd$ at NLO level in the conventional model II and found that the NLO
corrections enhance the LO results by about $18\%$.
In Ref.\cite{chao99}, the LO new physics
contributions to $\dmd$ have been calculated in model III and the constraint on the Yukawa
coupling $\ltt$ was also given by neglecting the large uncertainty of
the non-perturbative parameter $\fbbd$. We here will calculate
the charged Higgs contribution to mass splitting $\dmd$ at the NLO level in the model
III, use the new high precision data to constrain $|V_{td}|$ and $|\ltt|$
and consider the effects of the large uncertainty of $\fbbd$.

This paper is organized as follows. In Sec.~II, we describe the basic
structures  of the model III, give a brief review about the known constraints
on model III presented in previous works. In Sec.~III, we calculate the mass splitting
$\dmd$ at the NLO level in model III, and draw the constraint on $|\ltt|$ by employing
the new precision data. In Sec.~IV, the NLO new physics contributions to
the branching ratio $\brbxsga$ in model III are calculated and analyzed in great detail.
The numerical results for the conventional model II at NLO level are also presented as a
comparison. The conclusions are included in the final section.

\section{Structure of model III and constraint} \label{sec-2}

The simplest extension of the SM is the so-called two-Higgs-doublet models
\cite{2hdm,gunion90}. In such models, the tree
level flavor changing neutral currents(FCNC's) are absent if one
introduces an {\it ad hoc} discrete symmetry to constrain the 2HDM
scalar potential and Yukawa Lagrangian. Let us consider a Yukawa
Lagrangian of the form\cite{atwood97}
\beq
{\cal L}_Y &=& \eta^U_{ij}\bar{Q}_{i,L} \tilde{\phi_1}U_{j,R} +
\eta^D_{ij}\bar{Q}_{i,L} \phi_1 D_{j,R} +\xi^U_{ij}\bar{Q}_{i,L}
\tilde{\phi_2}U_{j,R} +\xi^D_{ij}\bar{Q}_{i,L} \phi_2 D_{j,R}+
H.c., \label{leff}
\eeq
where $\phi_{i}$ ($i=1,2$) are the two Higgs doublets, $\tilde{\phi}_{1,2}=
i\tau_2 \phi^*_{1,2}$, $Q_{i,L}$ ($U_{j,R}$) with $i=(1,2,3)$ are
the left-handed isodoublet quarks (right-handed   up-type quarks),
$D_{j,R}$  are the right-handed  isosinglet  down-type quarks,
while $\eta^{U,D}_{i,j}$  and $\xi^{U,D}_{i,j}$ ($i,j=1,2,3$ are
family index ) are generally the nondiagonal matrices of the
Yukawa coupling. By imposing the discrete symmetry
\beq
\phi_1 \to - \phi_1, \phi_2 \to \phi_2, D_i \to - D_i, U_i \to  \mp U_i
\eeq
one obtains the so called Model I and Model II. In Model I the
third and fourth term in eq.(\ref{leff}) will be dropped by the
discrete symmetry, therefore, both the up- and down-type quarks
get mass from Yukawa couplings to the same Higgs doublet $\phi_1$,
while the $\phi_2$ has no Yukawa couplings to the quarks. For
Model II, on the other hand, the first and fourth term in
Eq.(\ref{leff}) will be dropped by imposing the discrete symmetry.
Model II has, consequently, the up- and down-type quarks getting
mass from Yukawa couplings to two different scalar doublets
$\phi_1$ and $\phi_2$.

During past years, models I and II have been studied
extensively in literature at LO \cite{lo2hdm} and NLO level
\cite{bg98,ciuchini98,crs98,gm01}  and tested experimentally. The model
II has been very popular since it is the building block of the
minimal supersymmetric standard model. In this paper, we focus on
the third type of 2HDM \cite{hou92}, usually known as model
III \cite{hou92,atwood97}. In model III, no discrete symmetry
is imposed and both up- and down-type quarks then may have
diagonal and/or flavor changing couplings with $\phi_1$ and
$\phi_2$. As described in \cite{atwood97}, one can choose a
suitable basis $(H^0, H^1, H^2, H^\pm)$ to express two Higgs
doublets \cite{atwood97}
\beq
\phi_1 = \frac{1}{\sqrt{2}}\left (
\begin{array}{c} \sqrt{2} \chi^+ \\ v + H^0 + i \chi^0\\
\end{array} \right ),\ \ \phi_2=\frac{1}{\sqrt{2}}\left (
\begin{array}{c} \sqrt{2} H^+ \\ H^1 + i H^2 \\ \end{array} \right ), \label{phi12}
\eeq
and take their vacuum expectation values as
the form
\beq <\phi_1> &=& \left ( \begin{array}{c} 0 \\
v/\sqrt{2}\\ \end{array} \right ), \ \  <\phi_2>=0, \label{vev}
\eeq
where $v=(\sqrt{2}G_F)^{-1/2}=246GeV$. The transformation
relation between $(H^0,H^1,H^2)$ and the mass eigenstates
$(\overline{H}^0, h^0, A^0)$ can be found in \cite{atwood97}. The
$H^\pm$ are the physical charged Higgs boson, $H^0$ and $h^0$ are
the physical CP-even neutral Higgs boson and the $A^0$ is the
physical CP-odd neutral Higgs boson. After the rotation of quark
fields, the Yukawa Lagrangian of quarks are of the form
\cite{atwood97},
\beq
{\cal L}_Y^{III} = \eta^U_{ij}\bar{Q}_{i,L}
\tilde{\phi_1}U_{j,R} + \eta^D_{ij}\bar{Q}_{i,L} \phi_1 D_{j,R}
+\hat{\xi}^U_{ij}\bar{Q}_{i,L} \tilde{\phi_2}U_{j,R}
+\hat{\xi}^D_{ij}\bar{Q}_{i,L} \phi_2 D_{j,R} + H.c.,
\label{lag3}
\eeq
where $\eta^{U,D}_{ij}$ correspond to the diagonal mass
matrices of up- and down-type quarks, while the neutral and
charged flavor changing couplings will be \cite{atwood97}
\footnote{We make the same ansatz on the $\xi^{U,D}_{ij}$
couplings as the Ref.\cite{atwood97}. For more details about the
definition of $\hat{\xi}^{U,D}$ one can see Ref.\cite{atwood97}. }
\beq
\hat{\xi}^{U,D}_{neutral}= \xi^{U,D}, \ \ \hat{\xi}^{U}_{charged}=
\xi^{U}V_{CKM}, \ \ \hat{\xi}^{D}_{charged}= V_{CKM} \xi^{D},
\label{cxiud}
\eeq
with
\beq
\xi^{U,D}_{ij}=\frac{ g\,\sqrt{m_im_j}}{\sqrt{2}\mw } \lambda_{ij},
\label{lij}
\eeq
where $V_{CKM}$ is the Cabibbo-Kobayashi-Maskawa mixing matrix \cite{ckm}, $i,j=(1,2,3)$
are the generation index. It is easy to see from Eq.(\ref{lij}) that the
the FCNC within the first two generations are naturally suppressed by the small quark
masses, while a larger freedom in still allowed for the FCNC involving the top and bottom
quark. The coupling constants $\lambda_{ij}$ are  free parameters to be determined
by experiments, and they may also be complex.

In the conventional model I and model II, the only additional contribution to the
$B \to X_s \gamma$ decay  with respect to the SM comes from the charged Higgs boson-top
quark penguin diagrams and depends on the mass of
the charged Higgs boson,  $\mhp$,  and on $\tan{\beta} = v_2 / v_1$, where $v_{1,2} $
are the vacuum expectation values of $\phi_{1,2}$. From currently available studies
at NLO level \cite{bg98,ciuchini98,crs98,gm01},
one get to know the following main features of model II:
\begin{itemize}

\item
The charged Higgs penguins interfere constructively with  their SM counterparts,
and thus always enhance the branching ratio $\brbxsga$. The excellent agreement between the
theory and experiments for the decay rate  therefore leads to strong lower bound
on the mass $\mhp$. One typical lower bound ($99\% C.L.$) at NLO level
as given in Ref.\cite{gm01} is
\beq
\mhp > 315 GeV
\eeq
for any value of $\tan{\beta}$.

\item
The inclusion of NLO corrections shift the lower bound up by about $30\%$\cite{ciuchini98}.
In other words, the lower bound on $\mhp $ will become stronger in the NLO level
than that in the LO level. One  lower bound as given in Ref.\cite{ciuchini98} is
$\mhp > 258$ GeV using the LO calculation, but $\mhp > 368$ GeV using the NLO
calculation.

\item
The lower bound on $\mhp$ from the measured branching ratio $\brbxsga$ depends very
sensitively on small effects, and in particular on the way various errors are combined.
The difference is usually about $100$ GeV or even larger \cite{ciuchini98}.
Since the theoretical error is significantly reduced at the NLO level, improving the
calculation to the NLO has important effects on the lower bounds on $\mhp$.

\item
The bound on $\mhp$ from $B \to X_s \gamma $ decay is much stronger than
those from other experiments. As shown in Fig.4 of Ref.\cite{gm01}, for example,
the direct limit from LEP experiments is only $\mhp > 78.6$ GeV \cite{pdg2003}, the ratio
$R_b = \Gamma(Z \to b \bar{b})/\Gamma(Z\to hadrons)$ is relevant only for very small
$\tan{\beta}$, while rare $B \to \tau$ decays constrain $\mhp$ only for large $\tan{\beta}$.

\item
The $\tan{\beta}$ dependence of the lower bound saturates for
$\tan{\beta} \gtrsim 5$.

\end{itemize}

For the model I, however, no bound on $\mhp$ can be obtained from $B \to X_s \gamma$
\cite{gambino01}, since the charged Higgs loops interfere destructively with the SM penguin
diagrams  and decouple for large $\tan{\beta}$.

Although many phenomenological investigations  have been done in the framework of model III
\cite{hou92,atwood97,atwood96,aliev99,chao99,xiao00,xiao01}, the situation here is still not
as clear as the model II, since there are much more free parameters in model III than in model II.
As pointed in Ref.\cite{atwood97}, the data of $K^0-\bar{K}^0$ and $B_d^0-
\bar{B}_d^0$ mixing processes put severe constraint on the flavor changing
couplings involving the first generation of quarks. It is therefore reasonable to
assume that the Yukawa couplings involving the u and d quarks are zero:
$ \lambda_{uj}=\lambda_{dj}=0$ for $j=1,2,3$.

In Ref.\cite{chao99}, Chao et al., studied the decay
$\bsga$ at the leading order by assuming that only the couplings $\lambda_{tt}$ and
$\lambda_{bb}$ are non-zero. They found that the constraint on
$\mhp$ imposed by the CLEO data of $\bsga$ can be greatly relaxed
by considering the phase effects of $\lambda_{tt}$ and
$\lambda_{bb}$. The constraints from $B^0-\overline{B^0}$ mixing,
the neutron electric dipole moment(NEDM), the $Z^0$-pole parameter
$\rho$ and $R_b$ were also considered in Ref.\cite{chao99}.
The Chao-Cheung-Keung (CCK) scenario of model III \cite{chao99} has following advantages:
\begin{itemize}
\item
Since one keeps only the couplings $\lambda_{tt}$ and $\lambda_{bb}$
none zero, the neutral Higgs bosons do not contribute at tree
level or one-loop level. The new contributions therefore come only
from the charged Higgs loop diagrams with the heavy internal
top quark.

\item
The new operators $O_{9,10}$ and all flipped chirality partners of
operators $O_{1, \cdots,10}$ as defined in \cite{aliev99} do not
contribute to the decay $\bsga$.

\item
The free parameters in this model III are greatly reduced to
$\lambda_{tt}$, $\lambda_{bb}$ and $\mhp$.

\end{itemize}

In the following sections, we will calculate the NLO QCD contributions to the mass splitting
$\dmd$ and the branching ratio $\brbxsga$, to find the constraints on the parameters
$\ltt$, $\lbb$ and $\mh$ of model III. We will study the effects
of the NLO QCD contributions in detail, and will also compare the results in
model III with those in model II.

\section{$B^0- \bar{B}^0$ mixing  in model III }\label{sec-3}

\begin{figure}[htb]
\vspace{-120pt}
\begin{minipage}[]{\textwidth}
\centerline{\epsfxsize=\textwidth \epsffile{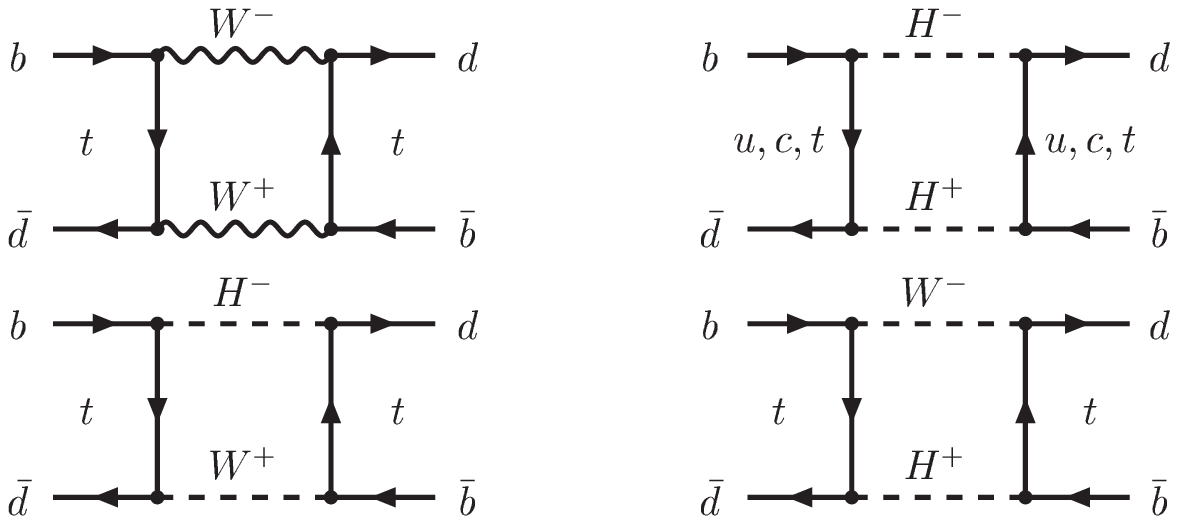}}
\vspace{-150pt}
\caption{ The box diagrams for the $B_d^0-\bar{B}_d^0 $ mixing in the framework of
2HDM at the lowest order.  More crossed diagrams are not shown. For those
Feynman  diagrams of NLO QCD corrections obtained by connecting any two quark
lines with a gluon line,  see the figures in Ref.\cite{urban98}. }
\label{fig:fig1}
\end{minipage}
\end{figure}

$B^0- \bar{B}^0$ mixing is in general a FCNC process generated through weak
interactions. At the lowest order of perturbation theory and in model III, the
corresponding box diagrams which generates this process is shown in
Fig.~\ref{fig:fig1}. The charged-Higgs boson contributions to $B^0- \bar{B}^0$ mixing
at leading order were calculated long time ago\cite{asw80}. The NLO QCD
corrections to $B^0- \bar{B}^0$ mixing was first presented in
Ref.\cite{buras90} for the SM, and in Ref.\cite{urban98} for the conventional 2HDM:
model I and model II. The possible constraints on model III from the measured parameter
$x_d = \Delta M_B /\Gamma_B$ were studied, for
example, in Refs.\cite{atwood97,grant95,chao99} at the LO level.

\subsection{The basic formulae}

The strengths of the $B_{q}^0 - \bar{B}_{q}^0$ mixing with $q\in (d,s)$  are
described by the mass
differences $\Delta M_{B_q}=M_H^{q} - M_L^q$ where the subscripts $H$ and $L$
denote the heavy and light mass eigenstates respectively.
The long distance contributions are estimated to be very small. The top
box diagram is strongly dominant, while the charm and mixed top-charm contributions
are entirely negligible.

Recently, great progress have been made in experimental measurements.
For $\dms$, a lower limit of
$\dms  > 14.1 ps^{-1}$ at $95\% C.L.$ is available. For $\dmd$,
however, it has been measured with high precision: the world average  \cite{hfag03,pdg2003}
is $\dmd = 0.502 \pm 0.007$ and dominated by the results of B factories.
At the end of the LEP-CDF-SLD era,
$\dmd$ has been measured with a relative precision of about $2.6\%$ \cite{ckm2003}.
After including the B factory measurements, the precision is now $1.2\%$ and high
enough to constrain the new physics contributions effectively.

On the theoretical side, the NLO theoretical prediction of $\dmd$ is available now in the
SM and in some new physics models beyond.
In SM the mass difference $\dmd$  can be written as \cite{buras96}
\beq
\dmd  = \frac{G_F^2}{6 \pi^2}  m_B |V_{td}|^2 (
\hat{B}_{B_d} f_{B_d}^2 ) \mw^2 \eta_B S_0(x_t)  ,
\label{eq:dmd}
\eeq
with
\beq
S_0(x) &=&\frac{4 x - 11 x^2 + x^3}{ 4 (1-x)^2} - \frac{3 x^3}{2
(1-x)^3}\ln[x], \label{eq:s0x}\\
\hat{B}_{B_d} &=&B_{B_d}(\mu) \left [\alpha_s^{(5)}(\mu)\right ]^{-6/23} \left
[ 1 + \frac{\alpha_s^{(5)}(\mu)}{4 \pi} \, J_5\right ],
\label{eq:bbd}\\
\eta_{B}&=& \left[\as(\mu_t)\right]^{6/23}
\times \left[1+\frac{\as(\mu_t)}{4\pi} \left(
 \frac{S_1(x_t)}{S_0(x_t)}+B_t - J_5 \right. \right. \non
&&\left. \left.   +\frac{\gamma^{(0)}}{2}
 \ln\frac{\mu^2_t}{M^2_W}+\gamma_{m0}
 \frac{\partial\ln S_0(x_t)}{\partial\ln x_t}\ln\frac{\mu^2_t}{M^2_W}
\right)\right] \label{eq:etab}
\eeq
where $x_t = \overline{m}_t(\mu_t) / \mw^2$
\footnote{$\overline{m}_q(\mu)$ is the running
q-quark mass in the modified minimal subtraction ($\overline{MS}$) scheme at the
renormalization scale $\mu$. For details, see Appendix \ref{app:input}.},
$\gamma^{(0)}=4$ and $\gamma_{m0}=8$ for $SU(3)_C$,
$B_t=17/3$ and $J_5=5165/3174$ in the NDR scheme\cite{buras96}, $m_B=5.279$ GeV \cite{pdg2003}
is $B_d^0$ meson mass, $\eta_B=0.55 \pm 0.01$ summarizes the NLO QCD
corrections \cite{buras90,urban98},  the function $S_0(x_t)$ describes the
dominant top-box contribution, $f_{B_d}$ is the $B_d^0$ meson decay constant,
and $\hat{B}$ is the renormalization group invariant and non-perturbative
parameter. There are a lot of works to estimate the values of $f_{B_d}$ and
$\hat{B}_{B_d}$ in lattice QCD calculation and in QCD sum rules\cite{ckm2003}.
The definitions of the various quantities in Eq.(\ref{eq:etab}) can be found
for example in Ref.\cite{buras96}. Using the input parameters as given in Appendix A,
we find numerically that $\eta_B=0.553$ and $0.496$ for $\mu_t = 170$ GeV or $\mu_t= \mw$,
respectively. The product $\eta_B S_0(x_t)$ has, however, a very weak
dependence on $\mu_t$ at NLO level: the uncertainty is only $0.3\%$ for $100$ GeV $\leq \mu_t \leq
300$ GeV.

With a well measured $\dmd$, one can determine $|V_{td}|$ from
Eq.(\ref{eq:dmd}) in the framework of the SM.
Using the input parameters as given in Appendix A, we find a limit on
$|V_{td}|$ from the measured $\dmd$ in Eq.(\ref{eq:dmd})
\beq
 |V_{td}| &=& \left [ 7.9 \pm 1.6 (\fbbd )
 \pm 0.2 ( \mt ) \pm 0.1 (  \eta_B  ) \right ] \times 10^{-3}\non
&=& (7.9 \pm 1.7 )\times 10^{-3}.\label{eq:vtdexp}
\eeq
where the uncertainties of $\fbbd$, $m_t$ and $\eta_B$ as listed in appendix A have been
considered and added in quadrature.
In the following calculations, we will use
this value of $|V_{td}|$ as input. It is easy  to see that
the error of $|V_{td}|$ is almost completely determined by the uncertainty of
the factor $f_{B_d} \sqrt{\hat{B}_{B_d}}$. If this uncertainty can be decreased by
a factor of two, we would find
\beq
 |V_{td}| &=& (  7.9 \pm 1.1 ) \times 10^{-3}. \label{eq:vtdexp-b}
\eeq

\begin{figure}[tb]
\vspace{3cm}
\centerline{\mbox{\epsfxsize=14cm\epsffile{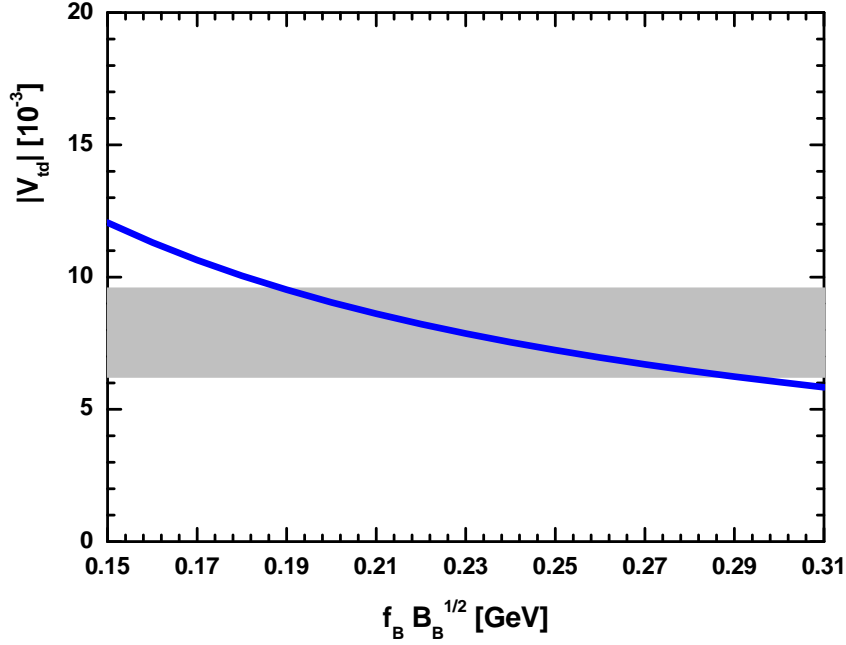}}}
\vspace{-4.5cm}
\caption{Contour plot of $|V_{td}|$ vs the non-perturbative parameter $\fbbd$ by using  the
well measured mass splitting  $\dmd=0.502 \pm 0.007 ps^{-1}$.
The shaded band corresponds to the
limit $|V_{td}|=(7.9 \pm 1.7)\times 10^{-3}$ obtained from the measured $\dmd$.
The width of the curve shows the effect of uncertainties of other quantities
in Eq.(\ref{eq:dmd}).}
\label{fig:fig2}
\end{figure}

Fig.\ref{fig:fig2} is the contour plot in $\fbbd-|V_{td}|$ plane obtained by the
using the data $\dmd=0.502 \pm 0.007 ps^{-1}$.  The shaded band in Fig.\ref{fig:fig2}
shows the allowed region as given inEq.(\ref{eq:vtdexp}). The solid  line shows
the $\fbbd$ dependence of $|V_{td}|$ and the width of the line shows the effect of
uncertainties of all other quantities appeared in Eq.(\ref{eq:dmd}).

\subsection{Mass splitting $\dmd$ in model II and III}

In the two Higgs doublet models, the charged Higgs boson contributes to the mass splitting
$\dmd$. In Ref.\cite{chao99}, the authors calculated the new physics
contribution to $\dmd$ in model III at the leading order and presented the constraint on the
$\lambda_{tt}-\mh$ plane by using the measured $x_d=\dmd/\Gamma_B$. But they did not
considered the effects of the large uncertainty of non-perturbative parameter $\fbbd$
and the new physics contribution to the parameter $\eta_B$.

Since $\dmd$ has been measured with very high precision, we use this quantity
directly instead of the parameter $x_d$ in our calculation. We will calculate
the charged Higgs contribution to mass splitting $\dmd$ at the NLO level by
extending the work in Ref.\cite{urban98} to the case of model III. We will
consider the effects of the large uncertainty of $\fbbd$.

In the framework of 2HDM, the NLO mass difference $\dmd$ can be written as  \cite{urban98}
\beq
\dmd = \frac{G_F^2}{6 \pi^2} m_B \mw^2 |V_{td}|^2( \hat{B}_{B_d} f_{B_d}^2 )
\eta_B(x_t,y_t) S_{2HDM}(x_t,y_t), \label{eq:dmd-m3}
\eeq
with
\beq
\eta_B(x_t,y_t)&=& \alpha_S(\mw)^{6/23} \left[ 1 + \frac{\alpha_S(\mw)}{4 \pi}
\left ( \frac{D_{2HDM}(x_t,y_t)}{S_{2HDM}(x_t,y_t)} - J_5 \right ) \right ],
\label{eq:etabm3}
\eeq
and
\beq
S_{2HDM}(x_t,y_t)&=& \left [  S_0(x_t) + S_{WH}(x_t,y_t) + S_{HH}(x_t,y_t)\right ] ,
\label{eq:s2hdm} \\
D_{2HDM}&=& D_{SM}(x_t) + D_H(x_t,y_t),\label{eq:d2hdm}
\eeq
where $x_t=\overline{m}^2_t(\mw)/\mw^2$ and $y_t=\overline{m}^2_t(\mw)/\mh^2$, and
the high energy matching scale has been chosen as $\mu=\mw$.
The functions $D_{SM}(x_t)$ and $D_H(x_t,y_t)$ in Eq.(\ref{eq:d2hdm}) describe the SM and
new physics part of the NLO QCD corrections to the mass splitting
$\dmd$ \cite{urban98}
\beq
D_{SM}(x_t) &=& C_F \left [ L^{(1,SM)}(x_t) + 3 S_0(x_t) \right ]
+ C_A  \left [ L^{(8,SM)}(x_t) +  5 S_0(x_t) \right ]\;,
  \label{eq:dsmxw}\\
D_{H}(x_t,y_t) &=& C_F \left [ L^{(1,H)}(x_t,y_t) + 3 \left (
S_{WH}(x_t,y_t) + S_{HH}(x_t,y_t) \right ) \right ] \non
& & + \;C_A \left [  L^{(8,H)}(x_t,y_t) + 5 \left ( S_{WH}(x_t,y_t)
+ S_{HH}(x_t,y_t) \right )  \right ]\,, \label{eq:dh}
\eeq
where $C_F=4/3$ and $C_A=1/3$ for $SU(3)_C$. The function $S_0(x_t)$ describes the
dominant top-box contribution in the SM and has been given in Eq.(\ref{eq:s0x}). The functions
$S_{WH}(x_t,y_t)$ and $S_{HH}(x_t,y_t)$ denote the new physics contributions
from the box diagrams with one or two charged Higgs involved\cite{urban98},
\beq
\nonumber
S_{WH}(x_t,y_t)&=& |Y|^2\,\frac{y_t\;x_t}4\left [\frac{(2 x_t -8 y_t)
                  \ln(y_t)}{(1-y_t)^2
                  (y_t-x_t)}+\frac{6x_t\ln(x_t)}{(1-x_t)^2(y_t-x_t)}\right.\\
  &&  \left. \ \ \ -\frac{8-2x_t}{(1-y_t)(1-x_t)}\right ]\; ,  \\
S_{HH}(y_t) &=& |Y|^4\,\frac{y_t\;x_t}4\left [ \frac{1+y_t}{(1-y_t)^2}+
                \frac{2y_t\ln[y_t]}{(1-y_t)^3}\right ]\; .
\eeq
And finally, the function $L^{(i,H)}$ ($i=1,8$)
describe the charged Higgs contribution \cite{urban98}
\beq
L^{(i,H)}(x_t,y_t) = 2 |Y|^2\,WH^{(i)}(x_t,y_t)\,+
2 |Y|^2\,
\Phi H^{(i)}(x_t,y_t)+\,  |Y|^4\,HH^{(i)}(y_t)\;.
\eeq
The explicit expressions of complicated functions $WH^{(i)}(x_t,y_t)$,
$\Phi H^{(i)}(x_t,y_t)$ and $HH^{(i)}(y_t)$ can be found in Ref.\cite{urban98}.

Following Ref.\cite{bg98}, we here  use the symbols $X$ and $Y$ to denote the
Yukawa couplings  between the charged Higgs boson and quarks in the general 2HDMs.
In the conventional model I and II,  the couplings $X$ and $Y$ are real
and given by
\beq
X&=& -\cot{\beta}, \ \ Y=\cot{\beta} \ \ {\rm (Model \ \  I)}\, , \\
X&=& \tan{\beta}, \ \ \ \ Y=\cot{\beta} \ \ \ {\rm (Model \ \ II)}\, .
\label{eq:xy-m2}
\eeq

In the Model III where only the couplings $\ltt$ and $\lbb$ are non-zero,
the relation between the couplings $(X,Y)$ and $(\ltt,\lbb)$ is simple
\beq
X= -\lbb, \ \ Y=\ltt \ \ {\rm (Model \ \ III)}\, .
\label{eq:xy-m3}
\eeq

By using the input parameters as given in Appendix A, the SM prediction for
$\dmd$ is
\beq
\dmd = 0.506^{+0.198}_{-0.160} \, ps^{-1} \, , \label{eq:dmd-sm}
\eeq
where the error comes from the uncertainty of parameter $\fbbd$.

\begin{figure}[tb]
\vspace{3cm}
\centerline{\mbox{\epsfxsize=14cm\epsffile{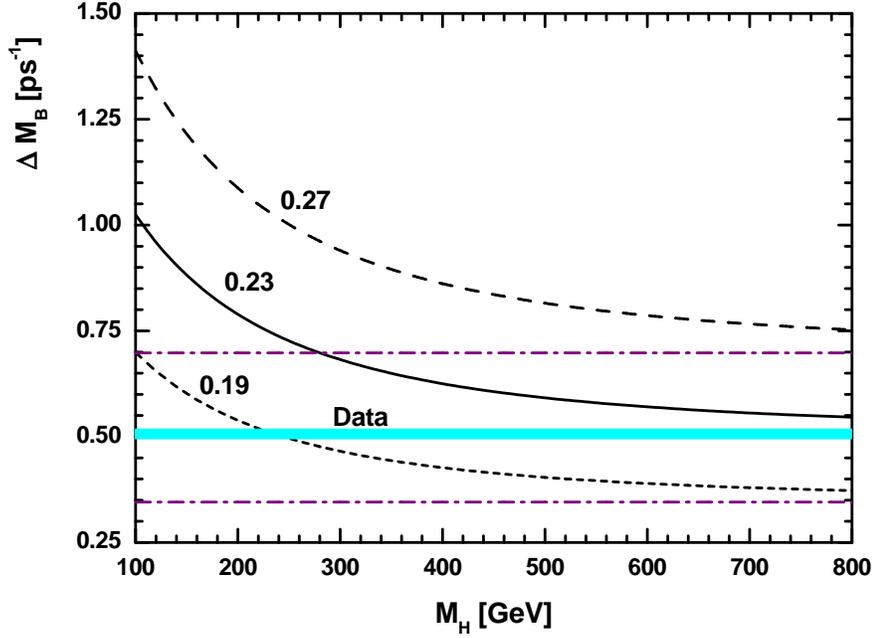}}}
\vspace{-4.5cm}
\caption{ Plots of the mass splitting $\dmd$ vs charged Higgs boson mass $\mh$ in
 model III.  }
\label{fig:fig3}
\end{figure}

In Fig.\ref{fig:fig3}, we show the  $\mh$ dependence of $\dmd$ in the model III
, assuming $\lambda_{tt}=1$.
The region between two horizontal dot-dashed lines corresponds to the SM
prediction as given in Eq.(\ref{eq:dmd-sm}).
The shaded horizontal band shows the world average $\dmd = 0.502 \pm 0.007 ps^{-1}$, and
its width corresponds to the error. The short-dashed, solid and dashed curves
in this figure show the model III predictions for $\fbbd=0.19,0.23$ and $0.27$,
respectively.

\begin{figure}[tb]
\vspace{3cm}
\centerline{\mbox{\epsfxsize=14cm\epsffile{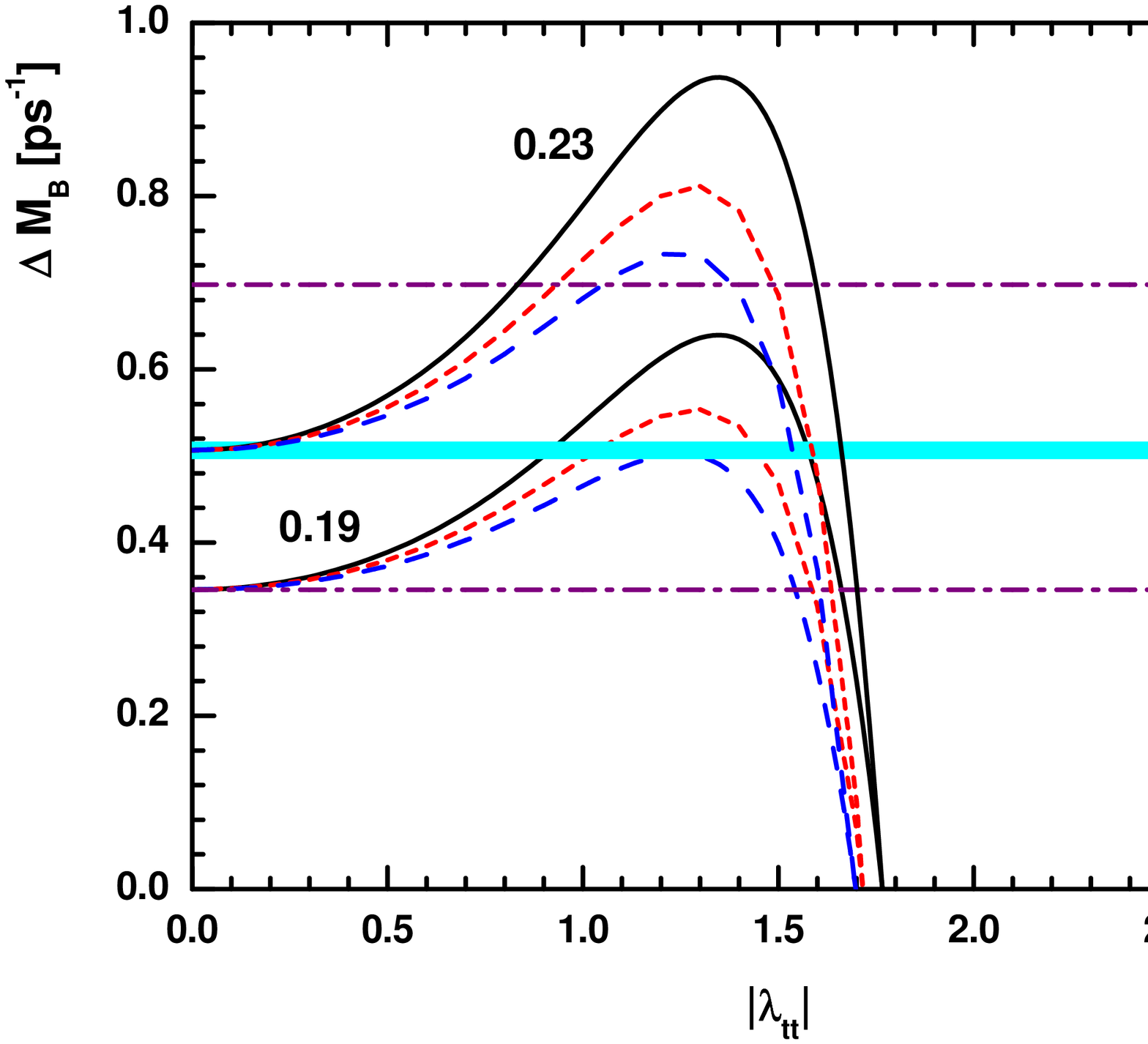}}}
\vspace{-4.5cm}
\caption{ Plots of the mass splitting $\dmd$ vs $|\ltt|$ in model III
for $\fbbd=0.19$ (the lower three curves) and $\fbbd=0.23$ (the upper three curves) and
for $\mh= 200$ (solid curves), $250$ (short-dashed curves) and $300$ GeV (dashed curves),
respectively.}
\label{fig:fig4}
\end{figure}

\begin{figure}[tb]
\vspace{3cm}
\centerline{\mbox{\epsfxsize=14cm\epsffile{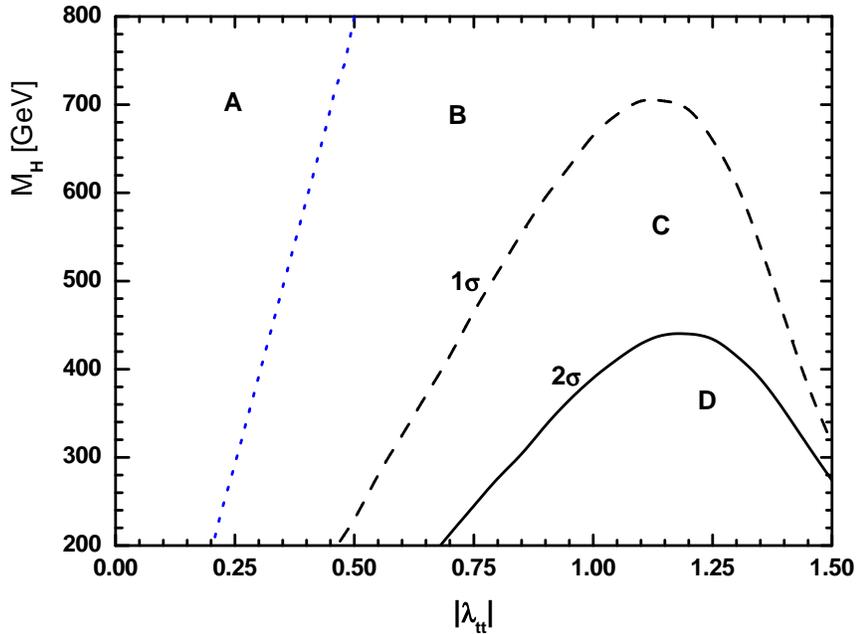}}}
\vspace{-4.5cm}
\caption{ Contour plot of the mass splitting $\dmd$ in the $\lambda_{tt}-\mh$ plane.
The measured $\dmd =0.502 \pm 0.007$ is used. The A area is allowed by considering the error
of $\dmd$ only, but both A and B area will be allowed if the uncertainty of $\fbbd$ is also
taken into account.}
\label{fig:fig5}
\end{figure}

From the well measured physical observable $\dmd$, one can find the constraint on
$|\ltt|$ in model  III.
In Fig.\ref{fig:fig4}, we show the  $|\ltt|$ dependence of $\dmd$ in the model III
for $\fbbd=0.19$ (the lower three curves) and $\fbbd=0.23$ (the upper three curves) and
for $\mh= 200$ (solid curves), $250$ (short-dashed curves) and $300$ GeV (dashed curves),
respectively. The shaded band and the region between two horizontal lines are
the same as in Fig.\ref{fig:fig3}.  From this figure, an upper bound on
$|\ltt|$ can be read off
\beq
|\ltt| \leq  1.7. \label{eq:ul-ltt}
\eeq
This bound is complementary to the constraint obtained from \fig{fig:fig5}.

Fig.\ref{fig:fig5} is the contour plot of the mass splitting $\dmd$ in the
$|\ltt|-\mh$ plane, where the measured $\dmd$ and
the relevant input parameters as given in appendix A have been used.
The area A in \fig{fig:fig5} will be allowed by the measured $\dmd$  within $2\sigma$
errors ( i.e., $\dmd=  0.502 \pm 0.014 ps^{-1}$),
if we do not consider the effect of the uncertainty of $\fbbd$ in drawing
this contour plot. This region corresponds to the allowed region in Fig.2 of Ref.\cite{chao99},
but much narrow than that one because of the great
progress of the experimental measurement of $\dmd$.
If we consider the effect of the large uncertainty of $\fbbd =0.23 \pm 0.04$ in our
calculations, the areas A plus B in the $|\ltt|-\mh$ plane will be allowed by
the measured $\dmd$ within $1\sigma$ error, while the areas A, B and C will be
allowed by the measured $\dmd$ within $2\sigma$ errors.

Since the large uncertainty of $\fbbd$ dominate the contour plot, one should take it
into account in the effort to limit the free parameters in model III. As discussed
previously, the uncertainty of $|V_{td}|$ are strongly correlated with the uncertainty of
$\fbbd$, we therefore consider the uncertainty of $\fbbd$ but use the central value
$|V_{td}|=0.0079$ in our calculation.

For the sake of experimental searches, one prefers a relatively light charged Higgs
boson. From the contour plot Fig.\ref{fig:fig5},  the region of $|\ltt| \gtrsim 0.7$
is disfavored if we expect existence of a light charged Higgs boson, while
the parameter space of
\beq
|\ltt| \approx 0.5 \ \ {\rm and} \ \ \mh \approx 200 {\rm GeV}, \label{eq:lttmh}
\eeq
is certainly allowed by the measured mass splitting $\dmd$.
The allowed region in the contour plot
will become narrow along with further improvement of the data and reduction of
the large theoretical uncertainty of parameter $\fbbd$ .

\begin{figure}[tb] 
\vspace{3cm}
\centerline{\mbox{\epsfxsize=14cm\epsffile{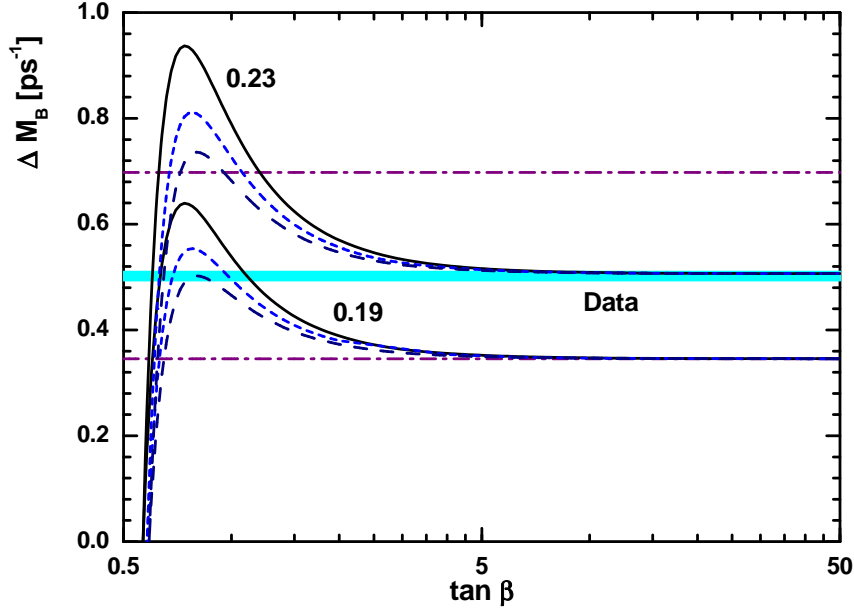}}}
\vspace{-4.5cm}
\caption{ Plots of the mass splitting $\dmd$ vs $\tan{\beta}$ in model II
for $\fbbd=0.19$ (the lower three curves) and $\fbbd=0.23$ (the upper three curves) and
for $\mh= 200$ (solid curves), $250$ (short-dashed curves) and $300$ GeV (dashed curves),
respectively. For details, see text.}
\label{fig:fig6}
\end{figure}

Since the new physics contributions to $\dmd$ depend on $|Y|^2$ and $|Y|^4$ only,
the charged Higgs contributions in the model II
and model III  will be the the same if we use the same value of the Yakawa coupling $|Y|$
as input. In model II, we have  $Y=1/\tan{\beta}$. The upper limit on $|\ltt|$
as given in Eq.(\ref{eq:ul-ltt}) can be translated to a lower limit on
$\tan{\beta}$,
\beq
\tan{\beta} \geq 0.6,
\eeq
as can be seen directly from Fig.\ref{fig:fig6}, where the upper and lower three
curves corresponds to
$\fbbd=0.19$ and $0.23$, respectively. The solid, short-dashed and dashed curves
the model III prediction for $\mh= 200$, $250$ and $300$ GeV,
respectively. The shaded band and the region between two horizontal lines are
the same as in Fig.\ref{fig:fig3}. One can also see from this figure that the
new physics contribution become negligible for $\tan{\beta} \geq 5$.


\section{The decay $\bxsga $ in model III }\label{sec-4}

In the absence of new light degrees of freedom, the new physics contributions to the
rare decay $\bxsga$ will
manifest itself through the new contributions to the Wilson coefficients of the
same operators involved in the SM calculation, or the new operators absent in
the SM, such as operators with different chirality. The excellent agreement
between SM theory and experimental data leads to strong constraints on many new
physics models beyond the SM.

In this section, we calculate the branching ratio $\calb (B \to X_s \gamma)$ in model III.
Here, the operator basis in the SM and model III under study is same.
The NLO QCD corrections will be included in model III by extending the calculations in
Ref.\cite{bg98}.  As a comparison, we also give the
numerical results in the model II where it is necessary.

\subsection{Effective Hamiltonian and operator basis}

In the framework of the SM, if we only take into account operators up to dimension 6 and put
$m_s=0$, the effective Hamiltonian for $\bsga$ at the scale $\mu$ reads\cite{bg98}
\beq
{\cal H}_{eff} = - \frac{4 G_F}{\sqrt{2}} V_{ts}V_{tb}^* \sum_{i=1}^{8}
C_i(\mu)O_i(\mu). \label{eq:heff}
\eeq
The operator basis  introduced by Chetyrkin, Misiak, and M\"unz (CMM)
\footnote{There are two popular operator basis used in literature. The standard basis
was defined for example in Ref.\cite{buras96}.
The second one is the CMM basis , where the fully
anticommuting $\gamma_5$ in dimensional regularization are employed
\cite{cmm97}.} are given by

\beq
{\cal O}_1 &= & (\bar{s}_L \gamma_\mu T^a c_L)\, (\bar{c}_L \gamma^\mu T_a b_L),\\
{\cal O}_2 &= & (\bar{s}_L \gamma_\mu c_L)\, (\bar{c}_L \gamma^\mu b_L),
\eeq
\beq
{\cal O}_3 &= & (\bar{s}_L \gamma_\mu b_L) \sum_q (\bar{q} \gamma^\mu q),\\
{\cal O}_4 &= & (\bar{s}_L \gamma_\mu T^a b_L) \sum_q (\bar{q} \gamma^\mu T_a q),\\
{\cal O}_5 &= & (\bar{s}_L \gamma_\mu \gamma_\nu \gamma_\rho b_L) \sum_q
 (\bar{q} \gamma^\mu \gamma^\nu \gamma^\rho q),\\
{\cal O}_6 &= & (\bar{s}_L \gamma_\mu \gamma_\nu \gamma_\rho T^a b_L) \sum_q
 (\bar{q} \gamma^\mu \gamma^\nu \gamma^\rho T_a q),\\
{\cal O}_7 &= & \frac{e}{16\pi^2} \,{\overline m}_b(\mu) \,
 (\bar{s}_L \sigma^{\mu\nu} b_R) \, F_{\mu\nu}, \\
{\cal O}_8 &= &  \frac{g_s}{16\pi^2} \,{\overline m}_b(\mu) \,
 (\bar{s}_L \sigma^{\mu\nu} T^a b_R) \, G^a_{\mu\nu},
\label{eq:cmm-basis}
\eeq
where $T_a$ ($a=1,\ldots , 8$) stands for $SU(3)_c$ generators, $g_s$ and $e$ are the
strong and electromagnetic coupling constants, $L,R=(1\mp \gamma_5)/2$ for the left
and right-handed projection operators, $O_1$ and $O_2$ are current-current operators,
$O_3-O_6$ are the QCD penguin operators, and $O_7$ and $O_8$ are electromagnetic
and chromomagnetic penguin operators.
In Eq.(\ref{eq:cmm-basis}), ${\overline m}_b(\mu)$ is the running b-quark mass in
the modified minimal subtraction ($\overline{MS}$) scheme at the renormalization
scale $\mu$ (see appendix A for details).

\subsection{NLO Wilson coefficients at the scale $\muw$ and $\mu_b$}

To the first order in $\as$, the effective Wilson coefficients at the
scale $\mu_W = {\cal O}(\mw) $ can be written as \cite{bg98}
\beq
C^{\,{\rm eff}}_i(\mu_W) =  C^{0,\,{\rm eff}}_i(\mu_W)
             + \frac{\as(\mu_W)}{4\pi} C^{1,\,{\rm eff}}_i(\mu_W) \, .
\label{eq:wc-mw}
\eeq

The LO Wilson coefficients at the matching energy scale $\mw$
takes the form \cite{bg98},
\beq
 C^{0,\,{\rm eff}}_2(\muw)  & = &  1\, ,                        \\
 C^{0,\,{\rm eff}}_i(\muw)  & = &  0,  \hspace*{1cm} (i=1,3,4,5,6)\, , \\
 C^{0,\,{\rm eff}}_7(\muw)  & = &   C_{7,SM}^0(\mw)   + |Y|^2  \, C_{7,YY}^0(\mw)
  +        (XY^*) \, C_{7,XY}^0(\mw) \, , \label{eq:c70mw}  \\
 C^{0,\,{\rm eff}}_8(\muw)  & = &   C_{8,SM}^0(\mw )  + |Y|^2  \, C_{8,YY}^0(\mw)
  +        (XY^*) \, C_{8,XY}^0(\mw) \, ,
\label{eq:c80mw}
\eeq
with
\beq
 C_{7,SM}^0  & = & \frac{3x_{t}^3-2x_{t}^2}{4(x_{t}-1)^4}\ln{x_t}
                     +\frac{-8x_{t}^3-5x_{t}^2+7x_t}{24(x_t-1)^3},
  \label{eq:c70-sm}   \\
 C_{8,SM}^0  & = &\frac{-3x_{t}^2}{4(x_{t}-1)^4}\ln{x_t}
                     +\frac{-x_{t}^3+5x_{t}^2+2x_t}{8(x_t-1)^3}\, ,
 \label{eq:c80-sm}\\
 C_{7,YY}^0  & = & \frac{3y_{t}^3-2y_{t}^2}{12(y_{t}-1)^4}\ln{y_t}
                     +\frac{-8y_{t}^3-5y_{t}^2+7y_t}{72(y_t-1)^3}\, ,
 ,\label{eq:c70-yy} \\
 C_{7,XY}^0 & = & \frac{y_t}{12} \left[
 \frac{-5y_t^2+8y_t-3+(6y_t-4)\ln y_t}{(y_t-1)^3} \right]\, ,\label{eq:c70-xy}\\
 C_{8,YY}^0  & = & \frac{-3y_{t}^2}{12(y_{t}-1)^4}\ln{y_t}
                     +\frac{-y_{t}^3+5y_{t}^2+2y_t}{24(y_t-1)^3} \, ,\label{eq:c80-yy} \\
 C_{8,XY}^0 & = & \frac{y_t}{4}  \left[ \frac{-y_t^2+4y_t-3- 2 \ln y_t}{(y_t-1)^3} \right] \,,
\label{eq:c80-xy}
\eeq
where $x_t=m_t^2/\mw^2$, $y_t=m_t^2/\mh^2$, and these leading order functions have no explicit
$\muw$ dependence.

The NLO Wilson coefficients at the matching scale $\mu_W$ in model III can be written as \cite{bg98}
\beq
 C_1^{1,\,{\rm eff}}(\muw) & = & 15 + 6\ln\frac{\mu_W^2}{\mw^2}\, , \\
 C_4^{1,\,{\rm eff}}(\muw) & = & E_0 + \frac{2}{3} \ln\frac{\mu_W^2}{\mw^2}
    + |Y|^2 \, E_H \, , \label{eq:c41effmw} \\
 C_i^{1,\,{\rm eff}}(\muw) & = &   0 \hspace*{1.5truecm} (i=2,3,5,6)\, , \\
 C_7^{1,\,{\rm eff}}(\muw) & = & C_{7,\smallsm}^{1}(\muw)
  + |Y|^2 \, C_{7,\smallyy}^{1}(\muw)
  +        (XY^*) \, C_{7,\smallxy}^{1}(\muw), \label{eq:c71effmw}  \\
 C_8^{1,\,{\rm eff}}(\muw) & = & C_{8,\smallsm}^{1}(\muw)
  + |Y|^2 \, C_{8,\smallyy}^{1}(\muw)
  +       (XY^*) \, C_{8,\smallxy}^{1}(\muw) \,,
\label{eq:c81effmw}
\eeq
where for $i=7,8$ the functions on the right-hand side of
Eqs.(\ref{eq:c71effmw},\ref{eq:c81effmw}) are
\beq
C_{i,\smallsm}^{1}(\muw)& = &
  W_{i,\smallsm} + M_{i,\smallsm} \ln\frac{\muw^2}{\mw^2}
 + T_{i,\smallsm}\left( \ln\frac{m_t^2}{\muw^2} -\frac{4}{3} \right)\, , \\
C_{i,\smallyy}^{1}(\muw) & = &
  W_{i,\smallyy} + M_{i,\smallyy} \ln\frac{\muw^2}{\mh^2} \,+ T_{i,\smallyy}
  \left( \ln\frac{m_t^2}{\muw^2} - \frac{4}{3} \right) \, , \label{eq:ci-yy}\\
C_{i,\smallxy}^{1}(\muw) & = &
  W_{i,\smallxy} + M_{i,\smallxy} \ln\frac{\muw^2}{\mh^2}
                 \,+ T_{i,\smallxy} \left( \ln\frac{m_t^2}{\muw^2} - \frac{4}{3} \right) \,.
\label{eq:ci-xy}
\eeq
The explicit expressions of functions $W_{i,j}$, $M_{i,j}$ and $T_{i,j}$
($i=7,8$ and $j=SM,YY,XY$ ) can be found in Ref.\cite{bg98}, and also listed in
Appendix B for the convenience of the reader.

\begin{figure}[tb] 
\vspace{3cm}
\centerline{\mbox{\epsfxsize=14cm\epsffile{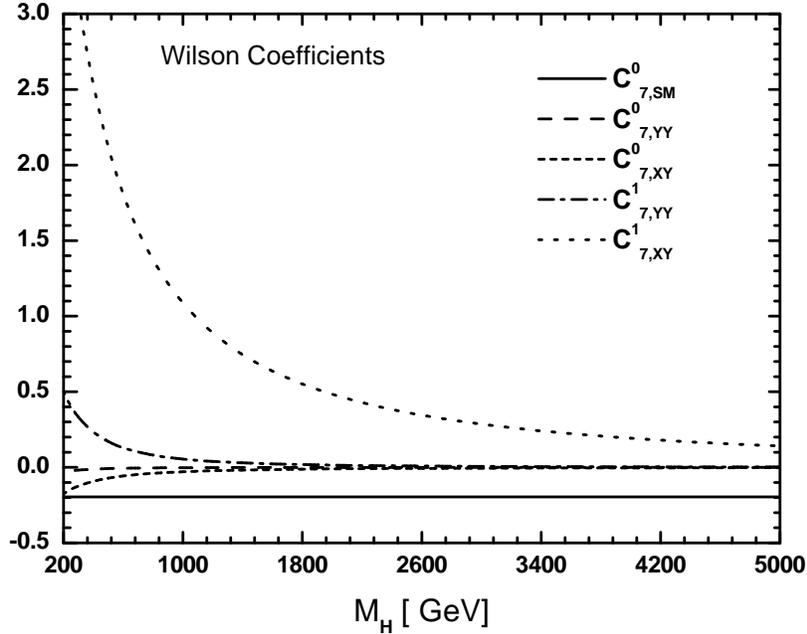}}}
\vspace{-4cm}
\caption{ Plots of $\mh$ dependence of  $C^0_{7,\smallsm}(\mw)$ (solid line)
and the four functions $C^{0,1}_{7,\smallyy}(\mw)$ (the dashed and dot-dashed curves )
and $C^{0,1}_{7,\smallxy}(\mw)$ (the short-dashed and dots curves). The
decoupling behaviour of new physics contributions can be seen clearly.}
\label{fig:fig7}
\end{figure}

The new physics contributions to the $B \to X_s \gamma$ decay are described by
the functions $C^{0,1}_{i,j}(\muw)$ with $i=(7,8)$ and $j=(YY,XY)$
as defined in Eqs.(\ref{eq:c70-yy}-\ref{eq:c80-xy}) and
(\ref{eq:ci-yy},\ref{eq:ci-xy}).
These eight functions depend on the unknown $\mh$ and the well measured $\mw$ and $m_t$ only,
and show clear decoupling behaviour when $\mh$ approaches infinity, as illustrated
in Fig.\ref{fig:fig7} for the four functions $C^{0,1}_{7,\smallyy}(\mw)$
and  $C^{0,1}_{7,\smallxy}(\mw)$.
Other four functions $C^{0,1}_{8,\smallyy}(\mw)$
and  $C^{0,1}_{8,\smallxy}(\mw)$ have the very similar decoupling behaviour.
Numerically, the LO new physics functions $C^{0}_{i,j}(\mw)$ are always negative and
have the same sign with their SM counterparts, while the NLO new physics
functions $C^{1}_{i,j}(\mw)$ are always positive
and have the opposite  sign with their SM counterparts.

Of course, the new physics contributions to $\bxsga$ also depend on the size and
sign of the couplings $X$ and $Y$, as can be seen easily from
Eqs.(\ref{eq:c70-yy}-\ref{eq:c80-xy}) and
(\ref{eq:ci-yy},\ref{eq:ci-xy}). For the conventional model II, we have
\beq
|Y|^2 = \frac{1}{\tan^2{\beta}}, \ \  X Y^* = 1, \label{eq:xy-m20}
\eeq
from the definition of $X$ and $Y$ as given in Eq.(\ref{eq:xy-m2}). For model III under study,
we have
\beq
|Y|^2 = |\ltt|^2, \ \  X Y^* = - |\ltt| |\lbb | e^{i \theta}, \label{eq:xy-m30}
\eeq
where $\theta = \theta_b-\theta_t $ is the relative phase of the coupling $\ltt$ and $\lbb$ in
model III. The sign of the second term in above equation can be negative or positive
depending on
the choice of $\theta$. In numerical calculations, we generally set
$\theta=0^\circ$, unless otherwise specified,
and will study the effects of a non-zero $\theta$ in the end of this section.

Since the heavy charged Higgs bosons have been integrated out at the scale
$\muw$, the QCD running of the the Wilson coefficients $C_i(\muw)$ down to the
lower energy scale $\mub = {\cal O}(m_b)$ after including the new physics
contributions is the same as in the SM.

For complete NLO analysis of the radiative decay $\bxsga$, only the Wilson coefficient
$C^{\,{\rm eff}}_7(\mub)$ should be known to NLO precision,
\beq
C^{\,{\rm eff}}_7(\mub) =  C^{0,\,{\rm eff}}_7(\mub)
             + \frac{\as(\mub)}{4\pi} C^{1,\,{\rm eff}}_7(\mub) \, ,
\label{eq:c7mub}
\eeq
with
\beq
 C^{0,\,{\rm eff}}_7(\mub)  & = &
  \eta^\frac{16}{23}  C^{0,\,{\rm eff}}_7(\muw)
 +\frac{8}{3} \left(\eta^\frac{14}{23} -\eta^\frac{16}{23}\right)
                   C^{0,\,{\rm eff}}_8(\muw)
 + \sum_{i=1}^8 h_i \,\eta^{a_i} \,C^{0,\,{\rm eff}}_2(\muw) \,,
\label{eq:c70mb}\\
 C^{1,\,{\rm eff}}_7(\mub) & = &
  \eta^{\frac{39}{23}} C^{1,\,{\rm eff}}_7(\muw) +
 \frac{8}{3} \left( \eta^{\frac{37}{23}} - \eta^{\frac{39}{23}} \right)
      C^{1,\,{\rm eff}}_8(\muw) \non
&& +\left( \frac{297664}{14283}   \eta^{\frac{16}{23}}
       -\frac{7164416}{357075} \eta^{\frac{14}{23}}
       +\frac{256868}{14283}   \eta^{\frac{37}{23}}
       -\frac{6698884}{357075} \eta^{\frac{39}{23}}
 \right) C^{0,\,{\rm eff}}_8(\muw) \non
&&  +\, \frac{37208}{4761} \left(\eta^{\frac{39}{23}} -\eta^{\frac{16}{23}} \right)
          C^{0,\,{\rm eff}}_7(\muw)     \non
&& + \sum_{i=1}^8  \left[  e_i \,\eta \,C^{1,\,{\rm eff}}_4(\muw)
 + f_i + k_i \eta +  l_i \,\eta\, C^{1,\,{\rm eff}}_1(\muw) \right ] \eta^{a_i} \,,
\label{eq:c71mb}
\eeq
where the symbol $\eta$ is defined as $\eta=\alpha_s(\muw)/\alpha_s(\mub)$, and the
``magic numbers" $a_i$, $h_i$, $e_i$, $f_i$, $k_i$, and $l_i$ in
Eq.(\ref{eq:c71mb}) are listed in Table \ref{tab:magic}.

The remaining coefficients are only needed to LO precision,
\beq
 C^{0,\,{\rm eff}}_j(\mub) & = & \sum_{i=1}^8 h_{ji} \eta^{a_i}, \ \  for  \ \
 j=1,\ldots, 6\, , \label{eq:cj0mb}\\
 C^{0,\,{\rm eff}}_8(\mub) & = &   \eta^\frac{14}{23}  C^{0,\,{\rm eff}}_8(\muw)
 + \sum_{i=1}^8 \hbar_i\,\eta^{a_i}\, , \label{eq:c80mb}
\eeq
where the ``magic numbers " $\hbar_i$ and $h_{ji}$ are also listed in Table
\ref{tab:magic}.
Following Ref.\cite{bg98}, the small coefficients $C_3^{0,\,{\rm eff}}(\mub), \ldots ,
C_6^{0,\,{\rm eff}}(\mub)$ will be neglected in numerical calculations.

\begin{table}[tb]
\begin{center}
\caption{The "magic numbers" appeared in the calculations of the the Wilson coefficients
$C_i(\mu)$ in the rare decay $\bsga$.}
\label{tab:magic}
\begin{tabular}{ c c c c c c c c c} \hline \hline
$i$&1&2&3&4&5&6&7&8\\ \hline
$a_i$&$\frac{14}{23}$&$\frac{16}{23}$&$\frac{6}{23}$&$-\frac{12}{23}$&0.4086&-0.4230&-0.8994&0.1456\\
\hline
$h_{1i}$&0&0&1&-1&0&0&0&0\\
$h_{2i}$&0&0&$\frac{2}{3}$&$\frac{1}{3}$&0&0&0&0\\
$h_{3i}$&0&0&$\frac{2}{63}$&$-\frac{1}{27}$&-0.0659&0.0595&-0.0218&0.0335\\
$h_{4i}$&0&0&$\frac{1}{21}$&$\frac{1}{9}$&0.0237&-0.0173&-0.01336&-0.0136\\
$h_{5i}$&0&0&$-\frac{1}{126}$&$\frac{1}{108}$&0.0094&-0.01&0.001&-0.0017\\
$h_{6i}$&0&0&$-\frac{1}{84}$&$-\frac{1}{36}$&0.0108&0.0163&0.0103&0.0023\\
\hline
$e_i$&$\frac{4661194}{816831}$&$-\frac{8516}{2217}$&0&0&-1.9043&-0.1008&0.01216&0.0183\\
$f_i$&-17.3023& 8.5027&4.5508 & 0.7519& 2.0040& 0.7476&-0.5358& 0.0914\\
$k_i$&9.9372  &-7.4878&1.2688 &-0.2925&-2.2923&-0.1461& 0.1239& 0.0812\\
$l_i$&0.5784  &-0.3921&-0.1429& 0.0476&-0.1275& 0.0317& 0.0078&-0.0031\\ \hline
$h_{i}$&2.2996&-1.0880&$-\frac{3}{7}$&$-\frac{1}{14}$&-0.6494&-0.0380&-0.0185&-0.0057\\
$\hbar_{i}$&0.8623&0&$0$&$0$&-0.9135&0.0873&-0.0571&0.0209\\ \hline
$\gamma_{i7}^{0,{\rm eff } }$ & $-\frac{208 }{243}$ & $\frac{416 }{81}$ &$-\frac{176 }{81}$ &
$-\frac{152 }{243}$&$-\frac{6272 }{81}$ & $\frac{4624}{243}$&$\frac{32}{3}$ & $-\frac{32}{9}$ \\
\hline \hline
\end{tabular}
\end{center}
\end{table}

\subsection{Branching ratio $\brbxsga$ in the SM}

The branching ratio of the inclusive radiative decay
$\bxsga$ can be written as
\beq
\brbxsga_{LO} &=& {\cal B}_{SL}
\left | \frac{ V_{ts}^*V_{tb} }{V_{cb}}\right |^2  \frac{6\alpha_{em}}{\pi
f(z)}  \,  \left |C_7^{0,eff}(\mu_b)\right |^2 ,  \label{eq:br-lo}
\eeq
at the LO level, and
\beq
\brbxsga_{NLO} &=& {\cal B}_{SL}
\left | \frac{ V_{ts}^*V_{tb} }{V_{cb}}\right |^2  \frac{6\alpha_{em}}{\pi f(z)
\kappa(z)} \,  \left [ |\overline{D}|^2 + A + \Delta \right ]\, ,  \label{eq:br-nlo}
\eeq
at the NLO level, where ${\cal B}_{SL}=(10.64\pm 0.23)\%$ is the measured semileptonic
branching ratio
of B meson \cite{pdg2003}, $\alpha_{em}=1/137.036$ is the fine-structure constant
\footnote{Based on the analysis in Ref.\cite{cm98b}, it is more appropriate to use
$\alpha_{em}^{-1}=137.036$ instead of $\alpha_{em}^{-1}=130.3 \pm 2.3$ in the study of
$B \to X_s \gamma$ decay.},
$z=m_c^{pole}/m_b^{pole}=0.29 \pm 0.02$ is the ratio of the quark pole mass
\footnote{Another choice of $z$ is $z=\overline{m}_c(\mu)/m_b^{pole}=0.22\pm 0.04$,
but we do not consider this issue here. For more details about quark mass effects in
$B \to X_s \gamma$ decay, see Ref.\cite{gm01}.},
$f(z)$ and $\kappa(z)$ denote the phase space factor and the QCD
correction \cite{cm78} for the semileptonic B decay
\beq
f(z)&=& 1-8 z^2+ 8z^6-z^8-24z^4 \log(z), \\
\kappa(z)&=&1-\frac{ 2\alpha_s(\mu) }{ 3\pi }\left [ ( \pi^2-\frac{31}{4} )(1-z)^2
+\frac{3}{2} \right ] + \frac{\delta_{SL}^{NP}}{m_b^2},
\label{eq:kz}
\eeq
where $\delta_{SL}^{\smallnp}$ denotes the nonperturbative correction to the semileptonic
B meson  decay,
\beq
\delta_{SL}^{\smallnp} &=& \frac{\lambda_1}{2} + \frac{3}{2}\lambda_2
\left [ 1- 4 \frac{(1-z^2)^4}{f(z)}\right ]. \label{eq:dslnp}
\eeq

The term $\overline{D} $ in Eq.(\ref{eq:br-nlo}) corresponds to the subprocesses
$b \to s \gamma$ \cite{bg98}
\beq
\overline{D} &=&  C^{\rm eff}_7(\mub)  +  V(\mub)\,, \label{eq:dbar}
\eeq
where the NLO Wilson coefficient $C^{\rm eff}_7(\mub)$ have been given in Eq.(\ref{eq:c7mub}),
and the function $V(\mub)$ is defined as
\beq
  V(\mub)=\frac{\as(\mub)}{4 \pi}\left \{  \sum_{i=1}^8 C_i^{0,\,{\rm eff}}(\mub)
   \left[r_i+ \frac{1}{2}\gamma_{i7}^{0,\,{\rm eff}}
             \ln \frac{m^2_b}{\mu^2_b}    \right]
             - \frac{16}{3} C_7^{0,\,{\rm eff}}(\mub) \right \} \,,
             \label{eq:vmub}
\eeq
where  the functions $r_i$ $(i=1,\ldots, 8)$ are the virtual correction functions (see
Appendix D of Ref.\cite{bg98}), $\gamma_{i7}^{0,\,{\rm eff}}$ are the elements
of the anomalous dimension matrix which govern the evolution of the Wilson
coefficients from the matching scale $\muw$ to lower scale $\mub$. The values of
$\gamma_{i7}^{0,\,{\rm eff}}$  have been given in the last line of Table \ref{tab:magic}, for
details see Ref.\cite{buras96}. The LO Wilson coefficients in Eq.(\ref{eq:vmub})
have been given in previous subsections.

In Eq.(\ref{eq:br-nlo}), term $A$ is the the correction coming from the
bremsstrahlung process $b \to s\gamma g$ \cite{ag91}
\beq
A = \frac{\alpha_s(\mub)}{\pi} \sum_{i,j=1;i \le j}^8 \,
  {\rm Re} \left\{ C_i^{0,\,{\rm eff}}(\mub) \,
            \left[ C_j^{0,\,{\rm eff}}(\mub)\right]^*
 \, f_{ij} \right\} \, \label{eq:aa}.
\eeq
The coefficients $f_{ij}$ have been defined and computed in Refs.\cite{ag91,cmm97}. We
here use the explicit expressions of those relevant $f_{ij}$ as given in Appendix E of
Ref.\cite{bg98}.

In order to relate the quark decay rate to the actual hadronic process, the
nonperturbative corrections obtained with the method of the heavy-quark effective theory
(HQET) should be included. The term $\Delta^{\smallnp}$ in Eq.(\ref{eq:br-nlo}) and the term
$\delta_{SL}^{\smallnp}$ in Eq.(\ref{eq:kz}) denote these nonperturbative corrections
scale as $1/m_b^2$ and $1/m_c^2$ \cite{falk94,voloshin97},
\beq
\Delta &=& \frac{\delta_\gamma^{\smallnp}}{m_b^2} \left |C_7^{0,eff}(\mu_b)\right |^2
+ \frac{\delta_c^{\smallnp}}{m_c^2} {\rm Re}\left \{ \left [ C_7^{0,eff}(\mu_b)\right ]^*
\left [ C_2^{0,eff}(\mu_b)-\frac{1}{6}C_1^{0,eff}(\mu_b) \right ] \right \}
\eeq
with
\beq
\delta_{\gamma}^{\smallnp} &=& \frac{\lambda_1}{2} - \frac{9}{2}\lambda_2, \label{eq:dgnp}\\
\delta_{c}^{\smallnp} &=& - \frac{\lambda_2}{9}, \label{eq:dcnp}
\eeq
where $\lambda_2 = (m_{B^*}^2 - m_B^2)/4 = 0.12$ GeV$^2$ extracted from B meson
mass splitting, while the parameter $\lambda_1=0.5\, {\rm GeV}^2$ has large uncertainty but
the overall $\lambda_1$ dependence largely cancels in the decay rate with the $\lambda_1$ term
in $\delta_{SL}^{\smallnp}$ as can be seen from Eq.(\ref{eq:br-nlo}).

Now we are ready to present numerical results of the branching ratios in the SM and 2HDMs,
specifically in model III. For the numerical evaluations, unless otherwise specified,
we use the central value of input parameters given in Appendix \ref{app:input}.
For the values of the matching scale and low energy scale, we always take $\muw =
\mw$, and $m_b/2 \leq \mu_b \leq 2 m_b$.

Using Eq.(\ref{eq:br-nlo}),  we find the SM prediction of the branching ratio $\brbxsga$
\beq
\brbxsga^{SM}_{NLO} &=& \left [ 3.52 ^{+0.02}_{-0.16}(\mub) \pm 0.08({\cal B}_{SL})
^{+0.20}_{-0.16} (\frac{m_c}{m_b})\right. \non
&& \left.
\pm 0.13 (\alpha_s) \pm 0.05(\mt)
\pm 0.04 \left (\left |\frac{V_{ts}^*V_{tb}}{V_{cb}}\right|^2\right )\right ]
\times 10^{-4} \non
&=& ( 3.52 \pm 0.28)\times 10^{-4}, \label{eq:brsm}
\eeq
where the major sources of errors are shown explicitly, and the individual  errors are
added in quadrature. The central value
$3.52\times 10^{-4}$ is obtained by using the central values of input parameters listed
in Appendix A and setting $\muw=\mw$ and $\mub=m_b$. In this paper, we only consider
the effects of the uncertainties of those six parameters as specified in Eq.(\ref{eq:brsm}).

\subsection{Branching ratio $\brbxsga$ in model III}

In the model III considered in this paper, the branching ratio has been parameterized in terms of
three parameters $\ltt, \lbb$ and $\mh$. From the limit on $|\ltt|$ obtained from the measured
mass splitting $\dmd$, the magnitude of the coupling $\ltt$ should be smaller than 1 if we
require the charged Higgs boson is relatively light, as can be seen from Fig.\ref{fig:fig5}.
In this section, we always set $|\ltt|=0.5$, unless otherwise specified.

\begin{figure}[tb] 
\vspace{3cm}
\centerline{\mbox{\epsfxsize=14cm\epsffile{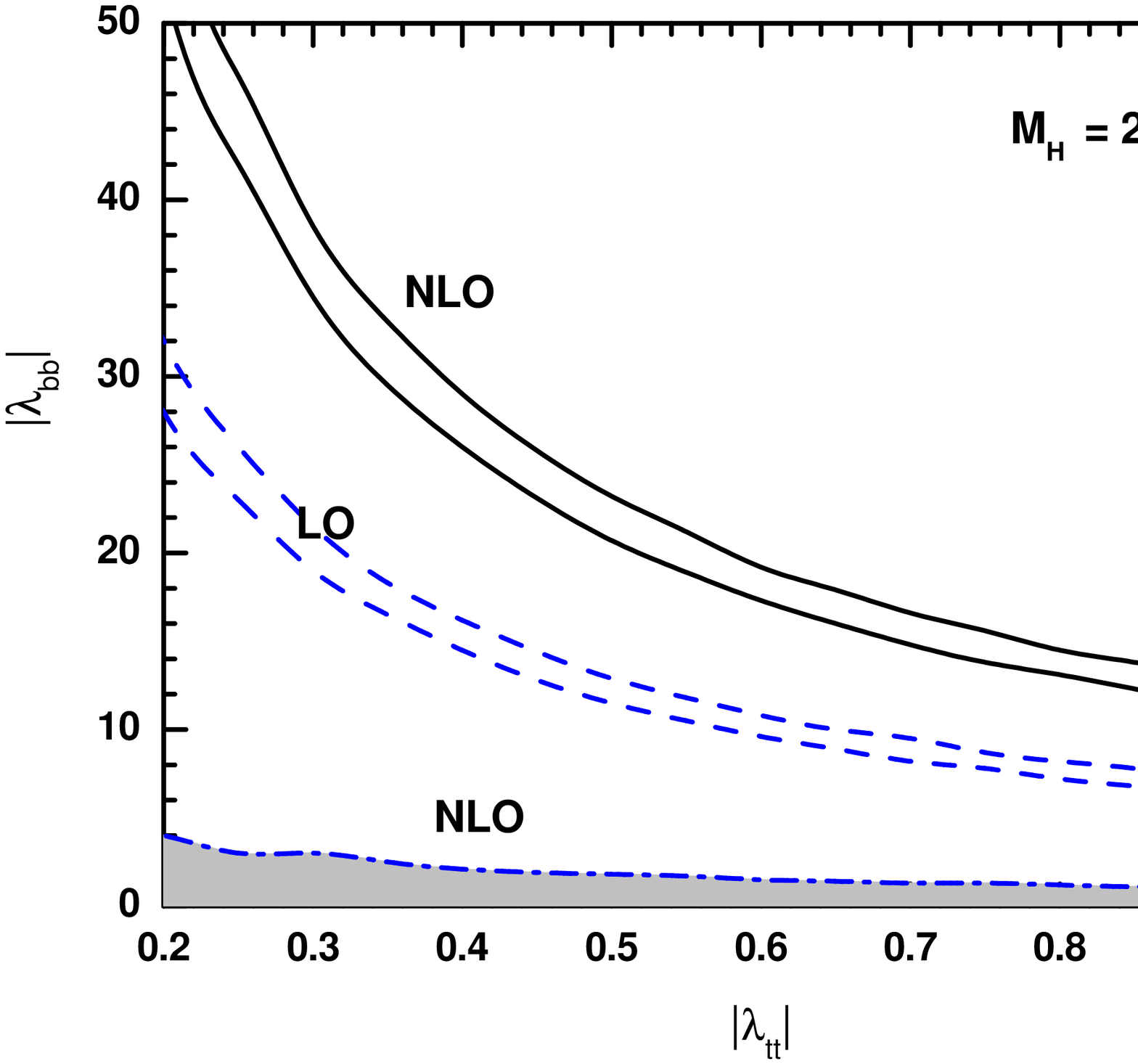}}}
\vspace{-4cm}
\caption{ Contour plot of the branching ratio $\brbxsga$ in the $|\ltt|-|\lbb|$ plane,
assuming $\mh=250$ GeV. }
\label{fig:fig8}
\end{figure}

We first check the common allowed regions in $|\ltt|-|\lbb|$ plane for given $\mh=250$ GeV.
Fig.~\ref{fig:fig8} is the contour plot in the $|\ltt|-|\lbb|$ plane by using the LO and NLO
theoretical predictions in model III and the measured branching ratio at the $2\sigma$ level:
$2.58\times 10^{-4} \leq \brbxsga \leq 4.1\times 10^{-4}$.
In this contour plot, the first allowed region is the shaded area close to the
X-axis where both $|\ltt|$ and $|\lbb|$ are all small, which is obtained when the NLO
theoretical prediction is employed.
The second allowed region is the area between two dashed curves
obtained when the LO theoretical prediction is employed. The third region is the region
between two solid curves allowed when the NLO theoretical prediction  is employed.

We choose one point of $(\ltt, \lbb )$ in each of three allowed regions of
Fig.\ref{fig:fig8} as typical choices
\beq
{\rm Case  A:}\ \ \  && (\ltt, \lbb) = (0.5,1), \label{eq:ltb-nlo1}\non
{\rm Case  B:}\ \ \  && (\ltt, \lbb) = (0.5,12), \label{eq:ltb-lo}\non
{\rm Case  C:}\ \ \  && (\ltt, \lbb) = (0.5,22), \label{eq:ltb-nlo2}
\eeq
which will be used as input in the following numerical calculations. Here the limit
on $|\ltt|$ from the measured $\dmd$ as studied in last section has been taken
into account.

Since the new physics contribution to the decay $B \to X_s \gamma$ is incorporated
through its correction to the Wilson coefficients $C^{0,1}_{7,8}(\muw)$,
we would like to check the size and sign of the new physics parts and their relative strength
to the SM part, to show the theoretical features of the NLO contributions and to
draw the constraint on $\mh$ by compare the theoretical predictions with the data.

\subsubsection{Case A}

We firstly consider the Case A: $(\ltt, \lbb) = (0.5,1)$.
Using the input parameters as given in Appendix A
and assuming $\mh=250$ GeV, we find the numerical results of $C_7^{0,1}$
\beq
 C^{0,\,{\rm eff}}_7(\mw)  & = & \begin{array}[t]{cccc}
  \underbrace{\ \ -0.1952\ \ }& \underbrace{\ \ -0.0057\ \ } &\underbrace{\ \ + 0.0730\ \ }
  & = -0.1280, \label{eq:ac70}\\
C^{0, \,{\rm eff}}_{7,\smallsm}(\mw) & |Y|^2 C^{0,\,{\rm eff}}_{7,\smallyy}(\mw)
& (XY^*) C^{0,\,{\rm eff}}_{7,\smallxy}(\mw)
& \\    \end{array}  \\
 C^{1,\,{\rm eff}}_7(\mw)  & = & \begin{array}[t]{cccc}
  \underbrace{\ \ -2.3712\ \ }& \underbrace{\ \ +0.1013 \ \ } &\underbrace{\ \  -1.7239\ \ } &
  = -3.9942, \label{eq:ac71}\\
C^{1,\,{\rm eff}}_{7,\smallsm}(\mw) & |Y|^2
C^{1,\,{\rm eff}}_{7,\smallyy}(\mw) & (XY^*) C^{1,\,{\rm eff}}_{7,\smallxy}(\mw)
& \\    \end{array}
\eeq
at the matching scale $\muw=\mw$, and
\beq
 C^{{\rm eff}}_7(m_b)  & = & \begin{array}[t]{cccc}
  \underbrace{\ \ -0.3137\ \ }& \underbrace{\ \ +0.0082\ \ } &
  \underbrace{\ \ + 0.0507\ \ } &  \underbrace{\ \ - 0.0142\ \ } \label{eq:ac7mb} \\
  C^{0,{\rm eff}}_{7,\smallsm}(m_b) & \Delta C^{1}_{7,\smallsm}(m_b)
&C^{0,\,{\rm eff}}_{7,\smallnp}(m_b) &\Delta C^{1}_{7,\smallnp}(m_b)
 \\    \end{array}\non
&=& -0.2690,
\eeq
at the lower scale $\mub=m_b$, where $\Delta C^{1}_{7,j}(\mu)=\frac{\as(\mu)}{4\pi}
C^{1, {\rm eff}}_{7,j}(\mu)$ with $j=({\rm SM, NP})$ denotes the NLO QCD corrections
to the corresponding Wilson coefficients.   It is easy to see that:
\begin{itemize}
\item
As shown in Eqs.(\ref{eq:ac70},\ref{eq:ac71}), the new physics contributions proportional
to $|Y|^2$ and $XY^*$ have the opposite sign and will cancel each other to some
degree. The net new physics contribution is relatively small and has the opposite sign with
its SM counterpart.

\item
The net NLO new physics contribution has the same sign and comparable in size with its SM
counterpart.

\item
At the low energy scale $\mub= {\cal O}(m_b)$,
the LO and NLO new physics contributions to $C_7^{\rm eff}(m_b)$
has the opposite sign, they will cancel each other. The total Wilson coefficient
$C_7^{\rm eff}(m_b)$ as given in Eq.(\ref{eq:ac7mb}) remains negative but changing from
$-0.3055$ to $-0.2690$ ( about a $12\%$ decrease in magnitude) after the cancellation
due to the new physics part.

\item
The new physics corrections are generally small in magnitude, since the couplings $X$
and $Y$ are all  relatively small for case A.

\end{itemize}

\begin{figure}[tb] 
\vspace{3cm}
\centerline{\mbox{\epsfxsize=14cm\epsffile{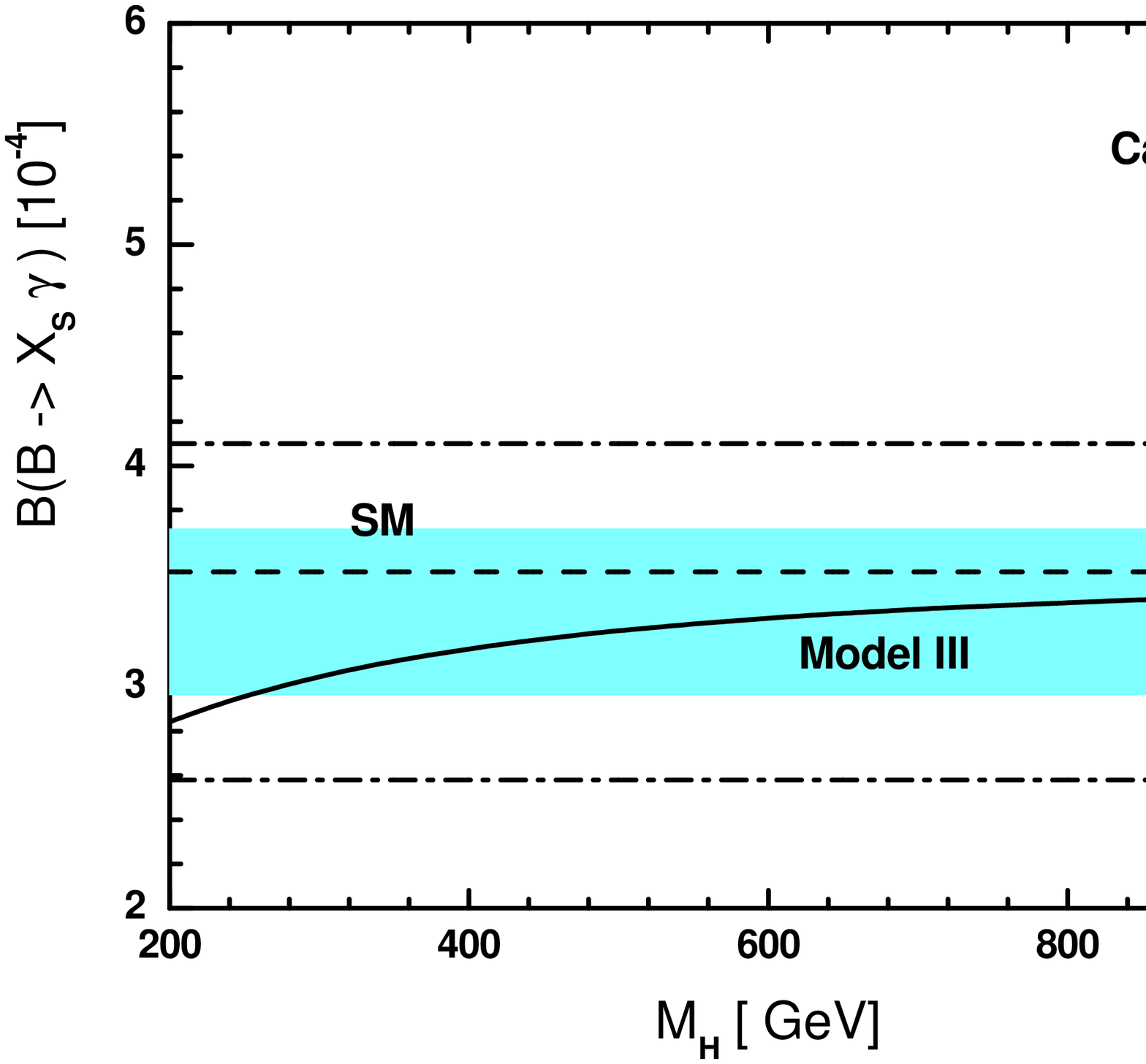}}}
\vspace{-4cm}
\caption{ The $\mh$ dependence of branching ratio $\brbxsga$ for the Case A: $(\ltt,\lbb)=(0.5,1)$
and at NLO level.
The band between two horizontal dot-dashed lines shows the data at $2\sigma$ level.
The dashed line and solid curve refers to the SM and Model III prediction, respectively.}
\label{fig:fig9}
\end{figure}

Explicit calculations also show that the new physics corrections to $V_{m_b}, A$ and
$\Delta $ terms are induced through the modification to the
Wilson coefficients $C^{0,\, {\rm eff}}_{7,8}(\mub)$. But for case A, such
corrections are small in size and therefore the theoretical prediction for
branching ratio $\brbxsga$ in this case agree well with the data within
$1\sigma$ error, as illustrated in Fig.\ref{fig:fig9}.
The region between two dot-dashed lines and the shaded part in Fig.\ref{fig:fig9} shows the
measured branching ratio with $2\sigma$ and $1\sigma$ error, respectively.
The dashed line  and solid curve shows the
central value of the SM and model III (Case-A) prediction, respectively.

Obviously, the theoretical prediction of the Case A agrees well with both the SM prediction
and the data because both $\ltt$ and $\lbb$ are small in size. But this case  is not
interesting theoretically, since the new physics effect is too small to be separated
from the SM contribution through experiments.

\subsubsection{Case B}

Now we turn to the case B: $(\ltt, \lbb) = (0.5,12)$.
Her only leading order contributions
in both SM and model III are taken into account, the branching ratio $\brbxsga$ completely
determined by the Wilson coefficient $C_7^{0,eff}(\mu_b)$ as shown in
Eq.(\ref{eq:br-lo}).

Using the input parameters as given in Appendix
\ref{app:input} and assuming $\mh=250$ GeV, we find numerically that
\beq
 C^{0,\,{\rm eff}}_7(\mw)  & = & \begin{array}[t]{cccc}
  \underbrace{\ \ -0.1952\ \ }& \underbrace{\ \ -0.0057\ \ } &\underbrace{\ \ + 0.8754\ \ }
  & = +0.6745, \\
C^{0,\,{\rm eff}}_{7,\smallsm}(\mw) & |Y|^2 C^{0,\,{\rm eff}}_{7,\smallyy}(\mw)
& (XY^*) C^{0,\,{\rm eff}}_{7,\smallxy}(\mw)
& \\    \end{array}
\label{eq:c70mw-b}
\eeq
at the matching scale $\muw=\mw$, and
\beq
 C^{0,\,{\rm eff}}_7(m_b)   =  \begin{array}[t]{ccc}
  \underbrace{\ \ -0.3137\ \ }& \underbrace{\ \ -0.0042\ \ } &
  \underbrace{\ \ + 0.6581\ \ } \\
  C^{0,\,{\rm eff}}_{7,\smallsm}(m_b) & C^{0,\,{\rm eff}}_{7,\smallyy}(m_b)
&C^{0,\,{\rm eff}}_{7,\smallxy}(m_b)  \\    \end{array} = +0.3402,
\label{eq:c70mb-b}
\eeq
at the lower scale $\mub=m_b$. It is easy too see from the numbers as given
in Eqs.(\ref{eq:c70mw-b},\ref{eq:c70mb-b}) that
\begin{itemize}
\item
The new physics contribution proportional to $XY^*$ term is much larger in size
than the one proportional to $X$.

\item
At both energy scales $\mw$ and $m_b$, the net new physics contributions to
$ C^{0,\,{\rm eff}}_7(\mu) $ are always positive and much larger in magnitude than its
SM counterpart. The total LO Wilson coefficient $ C^{0,\,{\rm eff}}_7(m_b)$
changed from $-0.3137$ in the SM to $+ 0.3402$ in model III.
\end{itemize}

From the $\bxsga$ decay, only the magnitude of $C_7(m_b)$ instead of its sign
can be constrained by the relevant data. The semileptonic decay $B \to K^{(*)} l^+ l^-$ is
sensitive to the sign of $C_7$, but the precision of the measurement is still not
high enough  to determine the sign of $C_7$.

\begin{figure}[tb] 
\vspace{3cm}
\centerline{\mbox{\epsfxsize=14cm\epsffile{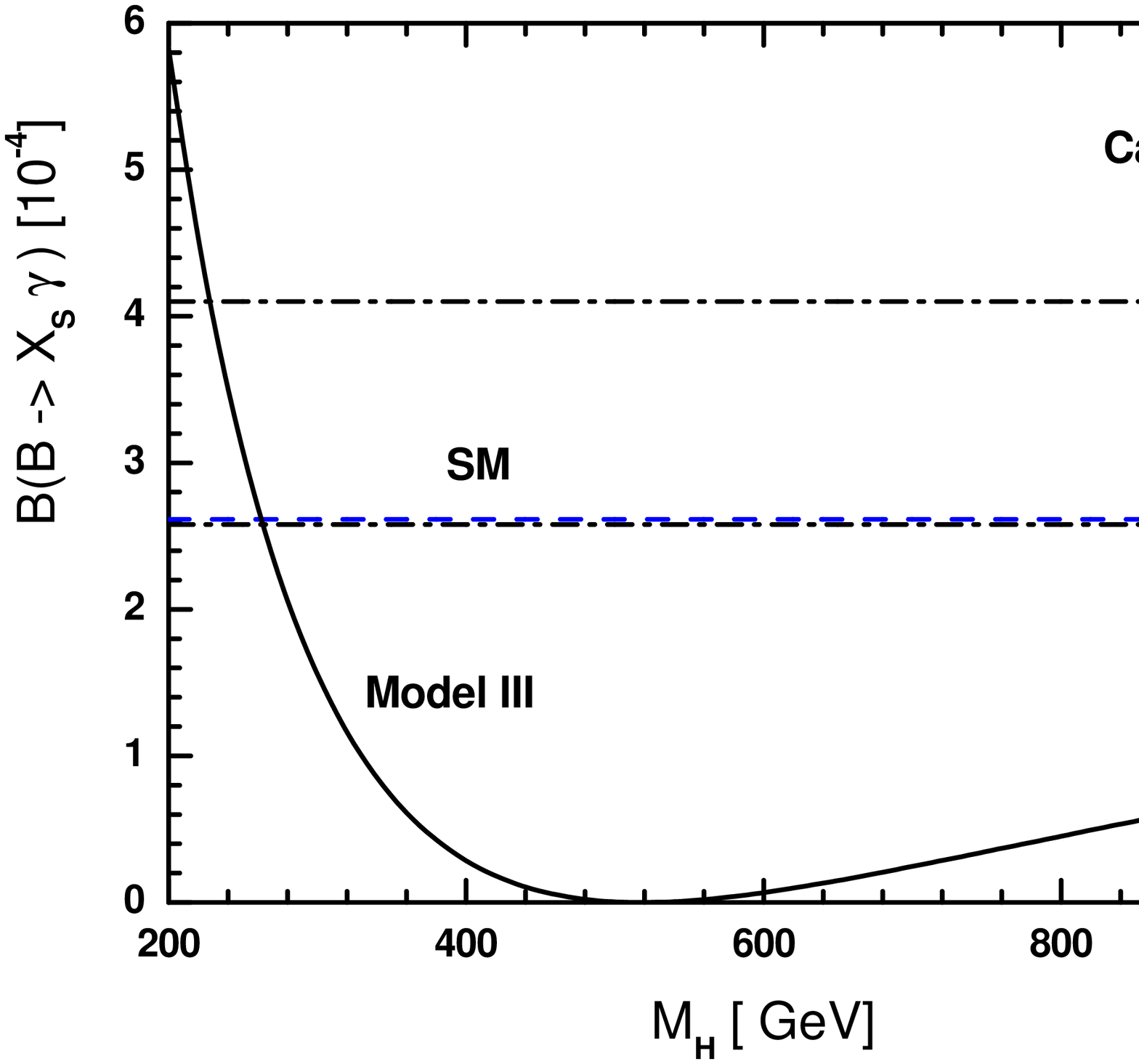}}}
\vspace{-4.5cm}
\caption{ The $\mh$ dependence of branching ratio $\brbxsga$ for the Case B:
$(\ltt,\lbb)=(0.5,12)$ and at LO level.
The band between two horizontal dot-dashed lines shows the data at $2\sigma$ level.
The dashed line and solid curve refers to the SM and Model III prediction, respectively.}
\label{fig:fig10}
\end{figure}

In Fig.~\ref{fig:fig10}, we show the $\mh$ dependence of the branching ratio
for the Case B. The dashed line shows the LO SM prediction $\brbxsga=2.61\times 10^{-4}$,
while the solid curve shows the model III prediction for Case B. It is interesting to note
that the LO SM prediction is marginally consistent with the data within $2\sigma$ errors,
but the NLO SM prediction agrees perfectly with the data.

One can also find the limit on $\mh$, $228 \leq \mh \leq 264$ GeV, from the
Fig.~\ref{fig:fig10} directly.
But one should know that the values of the lower and upper limits on $\mh$ will change
along with the variation of $\ltt$ and $\lbb$. The point here is that a
relatively  light charged Higgs boson is still allowed in model III even at the
leading order.

\subsubsection{Case C}

For Case C, the LO and NLO new physics contributions to the Wilson coefficients
$C_7(\mw)$ and $C_8(\mw)$ are rather large.
For given $(\ltt,\lbb)=(0.5,22)$ and assuming $\mh=250$
GeV, we find numerically that
\beq
C^{0,\,{\rm eff}}_7(\mw)  & = &  -0.1952\ \  + 1.5992 =  +1.4040, \\
C^{0,\,{\rm eff}}_8(\mw)  & = &  -0.0972\ \  + 1.4815 = +1.3843, \\
\frac{\alpha_s(\mw)}{4\pi} C^{1,\,{\rm eff}}_7(\mw)  & = &  -0.0229\ \  -0.3652 = -0.3881,  \\
\frac{\alpha_s(\mw)}{4\pi}C^{1,\,{\rm eff}}_8(\mw)  & = &  -0.0209\ \  - 0.2372= -0.2580,
\eeq
where the first and second term denote the SM and the new
physics contributions, respectively. Clearly, the new physics contributions are
always much larger than their SM counterparts in magnitude.

After the inclusion of new physics contributions, the
coefficients $C^{0,\,{\rm eff}}_7(\mw)$ and $C^{0,\,{\rm eff}}_8(\mw)$ become large and
positive. For NLO contributions, the new physics
parts are around ten times larger than the corresponding SM parts. Among the new physics parts,
the contribution proportional to term $XY^*$ is absolutely dominant over the one
proportional to $|Y|^2$, for example,
\beq
|\ltt|^2 C^0_{7,\smallyy}(\mw) = -0.0057, \ \
- |\ltt||\lbb|C^0_{7,\smallxy}(\mw)= 1.605
\eeq
for Case C and $\mh=250$ GeV, since $|\lbb|$ is now $40$ times larger than $|\ltt|$.

Put the things together, we find numerically that
\beq
 C^{{\rm eff}}_7(\mw)  & = & \begin{array}[t]{cccc}
  \underbrace{\ \ -0.1952\ \ }& \underbrace{\ \ -0.0411\ \ } &
  \underbrace{\ \ + 1.5992\ \ } &  \underbrace{\ \ - 0.6548\ \ }\\
  C^{0,\,{\rm eff}}_{7,\smallsm}(\mw) & \Delta C^{1}_{7,\smallsm}(\mw)
&C^{0,\,{\rm eff}}_{7,\smallnp}(\mw) &\Delta C^{1}_{7,\smallnp}(\mw)
 \\    \end{array}\non
&=& + 0.7081, \label{eq:c7mw-c}
\eeq
at the matching scale $\muw=\mw$, and
\beq
 C^{{\rm eff}}_7(m_b)  & = & \begin{array}[t]{cccc}
  \underbrace{\ \ -0.3137\ \ }& \underbrace{\ \ +0.0082\ \ } &
  \underbrace{\ \ + 1.2024\ \ } &  \underbrace{\ \ - 0.3320\ \ }\\
  C^{0,\,{\rm eff}}_{7,\smallsm}(m_b) & \Delta C^{1}_{7,\smallsm}(m_b)
&C^{0,\,{\rm eff}}_{7,\smallnp}(m_b) &\Delta C^{1}_{7,\smallnp}(m_b)
 \\    \end{array}\non
&=& +0.5649, \label{eq:c7mb-c}
\eeq
at the lower scale $\mub=m_b$.  From the numerical values in
Eqs.(\ref{eq:c7mw-c},\ref{eq:c7mb-c}), one can see that
\begin{itemize}
\item
In the SM, the NLO QCD contribution $ \Delta C^{1}_{7,\smallsm}(\mw)$ has the same sign
with its LO counterpart $ C^{0,\,{\rm eff}}_{7,\smallsm}(\mw)$ and is about $12\%$ of
$ C^{0,\,{\rm eff}}_{7,\smallsm}(\mw)$. After the QCD evolution from $\mw$ to
low energy scale $m_b$, the NLO QCD part
$\Delta C^{1}_{7,\smallsm}(m_b) $ changed its sign and tend  to cancel
$C^{0,\,{\rm eff}}_{7,\smallsm}(m_b)$. But the NLO QCD part is now only
$2.6\%$ of $ C^{0,\, {\rm eff}}_{7,\smallsm}(m_b)$,
and practically negligible for $\mub\approx m_b$.

\item
In model III and at the matching scale $\mw$, the NLO QCD  contribution
$ \Delta C^{1}_{7,\smallnp}(\mw)$ has
the opposite sign with its LO counterpart and
is as large as $41\%$ of $ C^{0,\,{\rm eff}}_{7,\smallnp}(\mw)$. Unlike the situation in the SM,
The NLO new physics contribution to $C_7$ in model III will cancel its LO counterpart
effectively, as shown explicitly in Eq.(\ref{eq:c7mw-c}).

\item
In model III and at the low energy scale $m_b$, the NLO QCD contribution
$ \Delta C^{1}_{7,\smallnp}(m_b)$ and $ C^{0,{\rm eff}}_{7,\smallnp}(m_b)$ still have
the opposite sign, but the ratio of these two parts is lowered from $41\%$ to
$28\%$.  The NLO and LO parts still cancel effectively.

\item
Through the QCD running from $\mw$ to $m_b$, the Wilson coefficient $C^{\rm eff}_{7,\smallsm}$
changed its value from $-0.2363$ to $-0.3055$, increased by about $29\%$ in magnitude. For
the new physics part, $C^{\rm eff}_{7,\smallnp}$ is decreased by $8\%$.

\end{itemize}

\begin{figure}[tb] 
\vspace{3cm}
\centerline{\mbox{\epsfxsize=14cm\epsffile{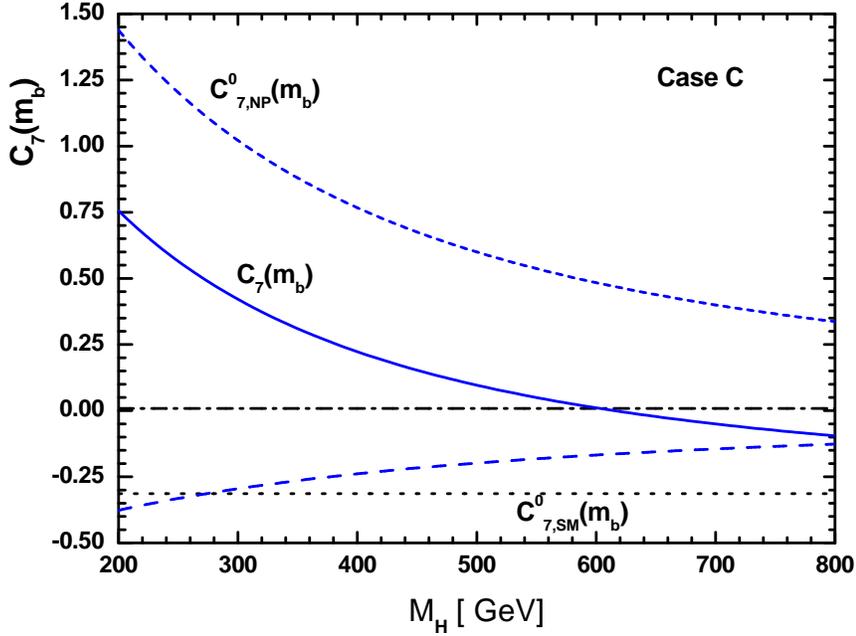}}}
\vspace{-4cm}
\caption{ The $\mh$ dependence of Wilson coefficient $C^{\rm eff}_{7}(m_b)$ and its SM and
model III parts for $(\ltt, \lbb)=(0.5,22)$ (i.e. the Case C).For details, see the text.}
\label{fig:fig11}
\end{figure}

The numerical values in Eqs.(\ref{eq:c7mw-c},\ref{eq:c7mb-c}) are obtained by
assuming $\mh=250$ GeV. For different values of $\mh$, the SM contributions remain unchanged,
but the new physics part as well as their sum $C^{\rm eff}_7(m_b)$ will change greatly,
as illustrated in Fig.~\ref{fig:fig11}. The horizontal dots and dot-dashed line  in
Fig.~\ref{fig:fig11} corresponds to $ C^{0,{\rm eff}}_{7,\smallsm}(m_b)=-0.3137$ and
$ \Delta C^{1,{\rm eff}}_{7,\smallsm}(m_b)=0.0082$, respectively.
The short-dashed and dashed curve shows the new physics contribution
$ C^{0,{\rm eff}}_{7,\smallnp}(m_b)$ and
$ \Delta C^{1}_{7,\smallnp}(m_b)$, respectively. Finally the solid curve is
the sum of those four parts. Since the dominant part $ C^{0,{\rm eff}}_{7,\smallnp}(m_b)$
is decreased rapidly for increasing $\mh$, the NLO part $ \Delta C^{1}_{7,\smallnp}(m_b)$
changed slowly and the SM parts remain unchanged, the Wilson coefficient
$C^{\rm eff}_7(m_b)$ is changing rapidly from large positive value to small negative value
within the range of $200 \leq \mh \leq 800$ GeV as shown by the solid curve
in Fig.~\ref{fig:fig11}, and will approach the value of its SM counterpart for
heavier charged Higgs boson.

One should note that when $C^{\rm eff}_7(\mub)$ becomes small, other previously ``small" NLO terms
$V(\mub)$, $A(\mub)$ and $\Delta$ may play an important rule. The typical
numerical values of those terms appeared in the curly brackets of Eq.(\ref{eq:br-nlo})
and their sum $R$,
\beq
R= |\overline{D}|^2 +A(m_b)+ \Delta,
\eeq
are listed in Table \ref{tab:ci-nlo}. In the SM, the
relative strength of individual terms are
\beq
&& C^{\rm eff}_7(m_b):V(m_b) = 1:0.079, \\
&& R: |\overline{D}|^2:A(m_b):\Delta = 1:0.98:0.030:(-0.011)
\label{eq:r-sm}
\eeq
The Wilson coefficient $ C^{\rm eff}_7(m_b)$ clearly dominate the total contribution
in the SM,
while the radiative correction $A(m_b)$ and the non-perturbative correction $\Delta$
play a minor rule since they are small in size and also cancel each other.

\begin{figure}[tb]
\vspace{3cm}
\centerline{\mbox{\epsfxsize=14cm\epsffile{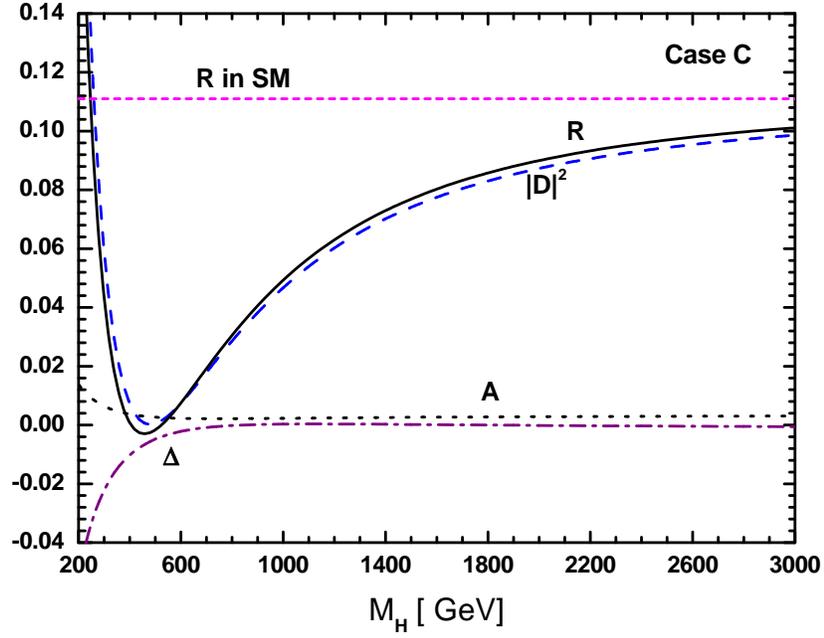}}}
\vspace{-4cm}
\caption{ The $\mh$ dependence of the terms appeared in the curly brackets of
Eq.(\ref{eq:br-nlo}) and the summation $R$ in the SM and model III of Case C.
For details, see the text.}
\label{fig:fig12}
\end{figure}

\begin{figure}[tbh] 
\vspace{3cm}
\centerline{\mbox{\epsfxsize=14cm\epsffile{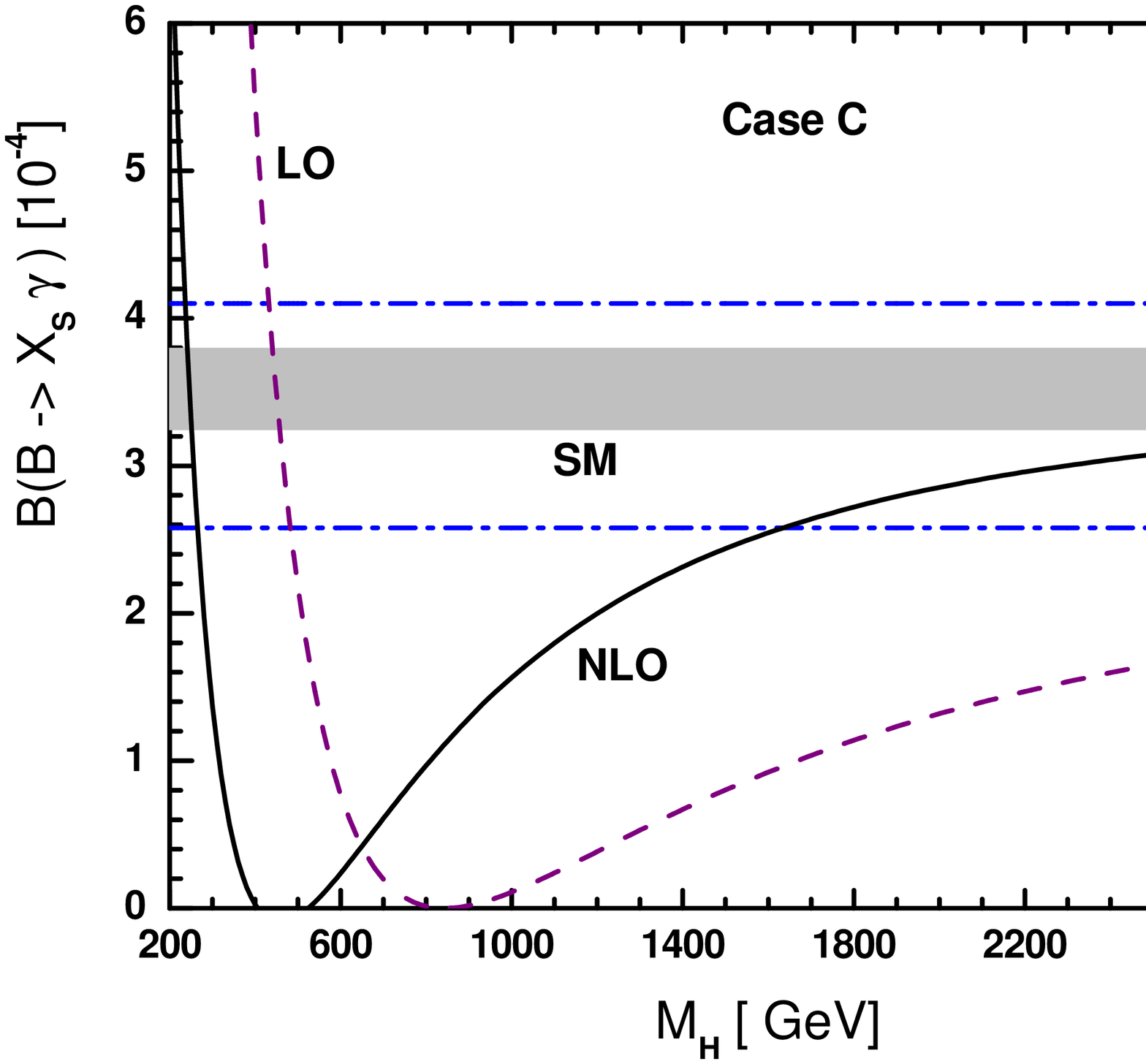}}}
\centerline{\mbox{\epsfxsize=14cm\epsffile{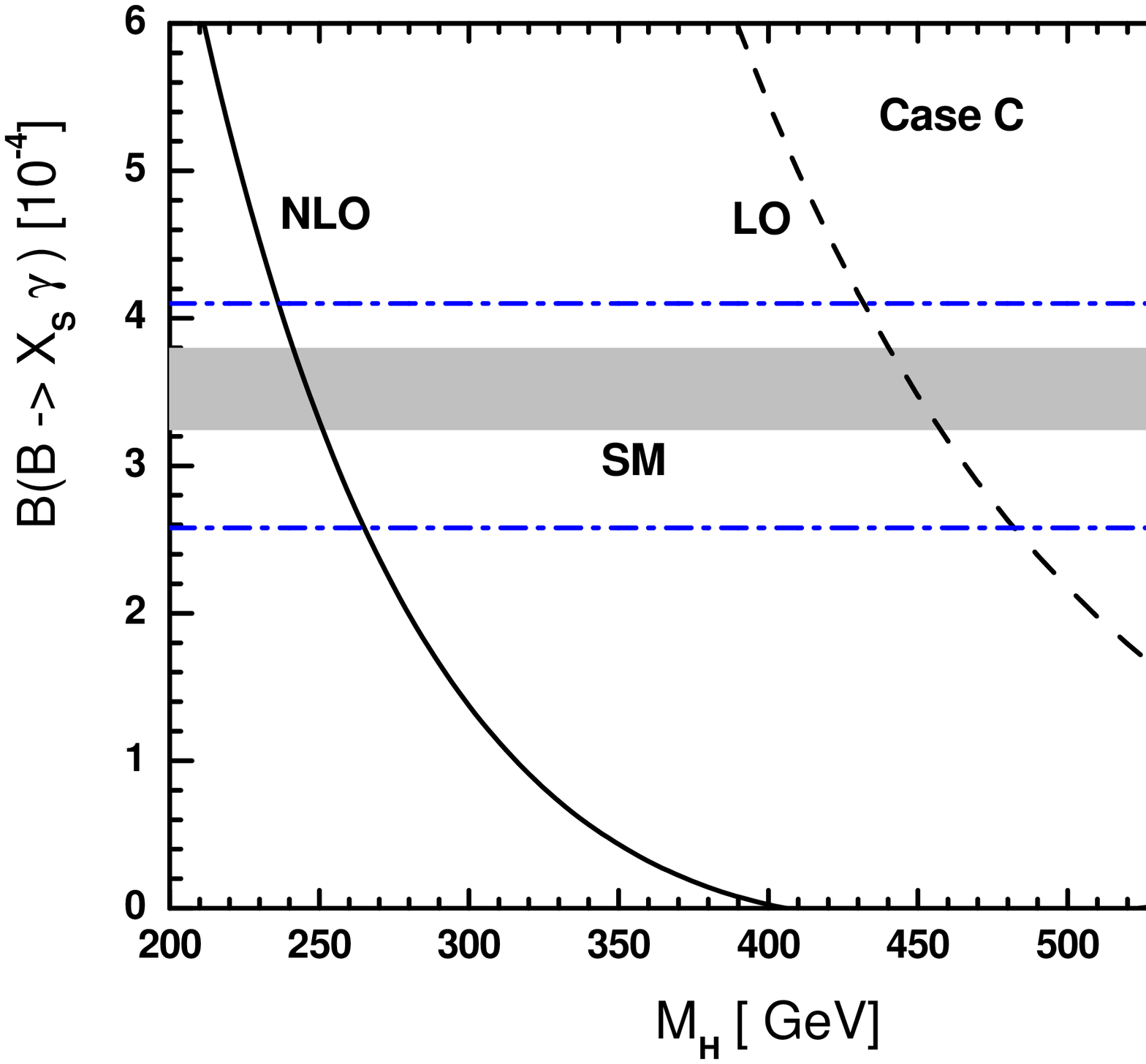}}}
\vspace{-2.5cm}
\caption{Plot of the branching ratio versus $\mh$ in model III of Case C.
The band between two horizontal lines shows the data with $2\sigma$ errors, the shaded band
shows the SM prediction: $\brbxsga =(3.52 \pm 0.28 )\times 10^{-4}$. The dashed and solid curve
shows the LO and NLO model III prediction, respectively. (b) is a magnification of the light
Higgs region of (a).  }
\label{fig:fig13}
\end{figure}

In model III, however, the situation is very different because $ C^{\rm eff}_7(m_b)$ and all
other terms appeared in curly bracket
of Eq.(\ref{eq:br-nlo}) will be changed by the inclusion of charged Higgs
contributions through the modified Wilson coefficients $ C^{0,\,\rm eff}_{7,8}(m_b)$ and
$ C^{1,\,\rm eff}_{7,8}(m_b)$, as can be seen from the numerical results listed in Table
\ref{tab:ci-nlo} and the curves shown in Fig.~\ref{fig:fig12}. The new
features are the following
\begin{itemize}
\item
Along with the increase of $\mh$, Wilson coefficient $ C^{\rm eff}_7(m_b)$
decrease rapidly, while $V(m_b)$ increase with less speed. Consequently, the `` dominant" term
$\dbars $ decreases more fast than $ C^{\rm eff}_7(m_b)$ does and approaches
the minimum of $\dbars=0.002$ for $\mh \approx 480$ GeV, which leads to a negative $R$ and
consequently a negative branching ratio: an
unphysical result. For more heavier charged Higgs boson, the new physics
contributions becomes smaller and smaller,
while the summation of individual terms in
the curly brackets of Eq.(\ref{eq:br-nlo}) restores to its SM value slowly, as illustrated by
the solid curve in Fig.~\ref{fig:fig12}. For $\mh > 600$ GeV, the terms $A(m_b)$
and $\Delta$ remain basically unchanged. As shown in Fig.\ref{fig:fig7},
the new physics contributions approaches zero when $\mh$ approaches infinity,
this is the so-called decoupling behavior of Higgs boson.

\item
From \tab{tab:ci-nlo} and \fig{fig:fig12}, one can infer that
there are two regions of $\mh$ allowed by the measured $\brbxsga$.
This point can be seen more directly from Fig.~\ref{fig:fig13}.

\end{itemize}

\begin{table}[tb]
\begin{center}
\caption{The numerical values of the NLO terms appeared in the curly brackets in
Eq.(\ref{eq:br-nlo}) and the branching ratio $\brbxsga$ (in unit $10^{-4}$) in the
SM and the  model III of Case C, for typical values of mass $\mh$ (in units of GeV).
The term $R$ is the summation of the terms $|\overline{D}|^2, A(m_b)$ and $\Delta$.}
\label{tab:ci-nlo}
\begin{tabular}{ c c |c c c c c c | c}\hline \hline
  & &$C^{\rm eff}_7(m_b)$ & $V(m_b)$&$|\overline{D}|^2$&$A(m_b)$&$\Delta$&$R$& ${\cal BR}$ \\
  \hline \hline
 SM&      &$-0.3055$& $-0.024 - 0.015 i$& $0.1089$&$0.0033$&$-0.0012$&$0.1110$&$3.52$ \\ \hline \hline
          &$200$ & $0.7559$& $-0.246 + 0.043 i$& $0.2615$&$0.0138$&$-0.0513$&$0.2240$&$7.10$\\
          &$250$ & $0.5649$& $-0.208 + 0.036 i$& $0.1283$&$0.0092$&$-0.0333$&$0.1042$&$3.30$\\
          &$300$ & $0.4210$& $-0.180 + 0.024 i$& $0.0591$&$0.0064$&$-0.0222$&$0.0433$&$1.37$\\
\cline{2-9}
          &$400$ & $0.2229$& $-0.140 + 0.020 i$& $0.0073$&$0.0037$&$-0.0102$&$0.0008$&$0.03$\\
Model III &$600$ & $0.0105$& $-0.096 + 0.009 i$& $0.0074$&$0.0023$&$-0.0022$&$0.0075$&$0.24$\\
          &$800$ &$-0.0951$& $-0.074 + 0.002 i$& $0.0286$&$0.0022$&$-0.0002$&$0.0305$&$0.97$\\
\cline{2-9}
         &$1000$ & $-0.1553$& $-0.061 -0.002 i$& $0.0467$&$0.0023$&$ 0.0003$&$0.0493$&$1.56$\\
         &$2000$ & $-0.2580$& $-0.037 -0.010 i$& $0.0873$&$0.0028$&$-0.0001$&$0.0900$&$2.85$\\
         &$3000$ & $-0.2827$& $-0.031 -0.012 i$& $0.0986$&$0.0031$&$-0.0006$&$0.1011$&$3.21$\\
\hline\hline
\end{tabular}
\end{center}
\end{table}

In \fig{fig:fig13}, we draw the $\mh$ dependence of the branching ratio
$\brbxsga$ in model III of Case C: $(\ltt,\lbb)=(0.5,22)$. In both (a) and (b),
the band between two dot-dashed horizontal lines shows the data within
$2\sigma$ errors. The shaded band shows the SM prediction and the error as given in
Eq.(\ref{eq:brsm}). The dash and solid curve corresponds to the LO and NLO model III
predictions, respectively. In order to show the effects of NLO corrections, we here
use the same values of $(\ltt,\lbb)=(0.5,22)$ as input for both LO and NLO theoretical
predictions. The Fig.\ref{fig:fig13}b is a magnification of the light Higgs
part of Fig.\ref{fig:fig13}a.
For given $\ltt=0.5, \lbb=22$ and $\mh = 250$ GeV,  we find numerically
\beq
\brbxsga^{III}_{NLO} &=& \left [ 3.48 ^{+0.18}_{-0.08}(\mub) \pm 0.08({\cal B}_{SL})
^{+0.42}_{-0.36} (\frac{m_c}{m_b})\right. \non
&& \left. ^{+0.52}_{-0.58} (\alpha_s) \pm 0.26(\mt)
\pm 0.04 \left (\left |\frac{V_{ts}^*V_{tb}}{V_{cb}}\right |^2\right )\right ]
\times 10^{-4} \non
&=& ( 3.48 \pm 0.74) \times 10^{-4}, \label{eq:brnp}
\eeq
where the central value of branching ratio is obtained by using the central values of input
parameters as given in appendix \ref{app:input}, while the six major errors from
the uncertainties of those input
parameters  are added in quadrature. For asymmetric errors we use the larger value
in making quadrature.

For Case C, the allowed regions of $\mh$ can be read off from \fig{fig:fig13} directly,
\beq
432\ \  \leq \mh \leq 478\ \  {\rm GeV}
\eeq
at the LO level, and
\beq
236\ \  \leq \mh \leq 266\ \  {\rm GeV}, \ \ {\rm and}
\ \ \mh \geq 1640 {\rm GeV}
\eeq
at the NLO level. The first allowed region, a region interested by many physicists
from the point of experimental searches,
is shifted to the lower part by about 200 GeV because of the inclusion of NLO contributions.

If we consider the effect of the theoretical error as given
in Eq.(\ref{eq:brnp}), the limits at the NLO level will become
\beq
226\ \  \leq \mh \leq 285\ \  {\rm GeV}, \ \ {\rm and}
\ \ \mh \geq 1120 {\rm GeV}.
\eeq
The first allowed region has a weak dependence on the theoretical error, but
the lower limit of the second allowed region of $\mh$ is very sensitive to the theoretical
error.

\begin{figure}[tbh]
\vspace{2.5cm}
\centerline{\mbox{\epsfxsize=14cm\epsffile{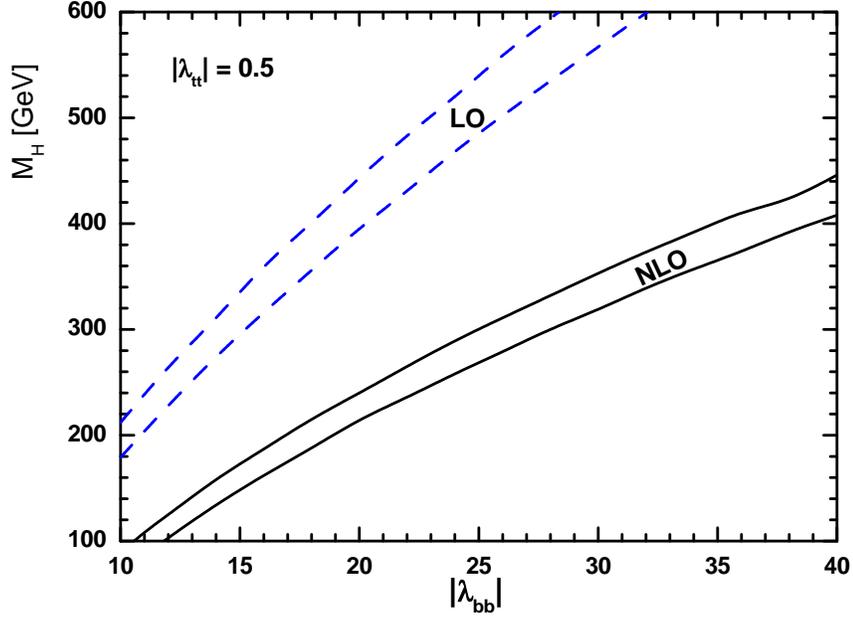}}}
\vspace{-4.5cm}
\caption{ Contour plot in $|\lbb|-\mh$ plane obtained by using the LO and NLO model
III predictions and the measured decay rate at $2\sigma$ level.
The regions between two dashed curves and two solid curves are allowed
when the LO and NLO theoretical predictions are employed.}
\label{fig:fig14}
\end{figure}

Fig.\ref{fig:fig14} is the contour plot in $|\lbb|-\mh$ plane obtained by using the LO
and NLO model III predictions and the measured decay rate at $2\sigma$ level, while assuming
$|\ltt|=0.5$. The regions between two dashed curves and two solid curves are allowed
when the LO and NLO theoretical predictions are employed. One can see from
Figs.\ref{fig:fig13} and \ref{fig:fig14} that a relatively light charged Higgs
boson in model III, say around or even less than 200 GeV, is still allowed by the measured
branching ratio $\brbxsga$.
For example, the region
\beq
 188 \leq \mh \leq 215 {\rm GeV}, \label{eq:limit}
\eeq
is allowed by the data if we set $(|\ltt|,|\lbb|)=(0.5,18)$ in model III.
This is a good news for future experimental searches!

\subsection{Comparison of the results in model II and III}

Between the conventional model II and the model III studied here, there is a
direct transformation. The results of model III can be reduced to the results
of model II by the substitution
\beq
\lbb =- X \to  -\tan{\beta} \ \ {\rm and} \ \ \ltt =Y  \to 1/\tan{\beta}.
\eeq

As a comparison, we here also present the LO and NLO model II prediction
for the branching ratio $\brbxsga$ and show the lower limits on $\mh$ obtained from the data.
For more details about the rare decay $\bxsga$ in the model II, one can see for
example Ref.\cite{bg98} and references therein.

By using the central values of input parameters as given in Appendix \ref{app:input},
and assuming $\tan{\beta}=4$, we find the numerical values of the NLO terms appeared
in the curly brackets in Eq.(\ref{eq:br-nlo}) and the branching ratio $\brbxsga$
in the model II for typical values of $\mh$, as listed in Table \ref{tab:ci-m2}.
In contrast to the model III of Case C, the differences between the values of the NLO
terms in the SM and model II are small and approaches zero when $\mh$ approaches infinity.
It is easy to see from Table \ref{tab:ci-m2} that the new physics correction
to the branching ratio becomes less than $6\%$ and $2\%$ for $\mh=1000$ and $2000$ GeV,
respectively.

If we consider the effects of uncertainties of input parameters, we find
the branching ratios at LO and NLO level
\beq
\brbxsga^{II}_{LO} &=& \left [ 3.61 ^{+0.74}_{-0.56}(\mub) \pm 0.08({\cal B}_{SL})
^{+0.32}_{-0.27} (\frac{m_c}{m_b})\right. \non
&& \left. \pm 0.09 (\alpha_s) \pm 0.08(\mt)
\pm 0.04 \left (\left |\frac{V_{ts}^*V_{tb}}{V_{cb}}\right |^2\right )\right ]  \times 10^{-4} \non
&=& ( 3.61 \pm 0.82 ) \times 10^{-4}, \label{eq:brm2lo}
\eeq
and
\beq
\brbxsga^{II}_{NLO} &=& \left [ 4.13 ^{+0.03}_{-0.17}(\mub) \pm 0.09({\cal B}_{SL})
^{+0.24}_{-0.20} (\frac{m_c}{m_b})\right. \non
&& \left. \pm 0.09 (\alpha_s) \pm 0.08(\mt)
\pm 0.04 \left (\left |\frac{V_{ts}^*V_{tb}}{V_{cb}}\right |^2\right )\right ]  \times 10^{-4} \non
&=& ( 4.13 \pm 0.34 ) \times 10^{-4}, \label{eq:brm2nlo}
\eeq
where the central value is obtained for $\mh=500$ GeV, and the errors connected with
the uncertainties of six input parameters are added in quadrature.

\begin{table}[tb]
\begin{center}
\caption{The numerical values of the NLO terms appeared in the curly brackets in
Eq.(\ref{eq:br-nlo}) and the branching ratio $\brbxsga$ (in unit $10^{-4}$) in the
model II with $\tan{\beta}=4$, for typical values of mass $\mh$ (in units of GeV).
The term $R$ is the summation of the terms $|\overline{D}|^2, A(m_b)$ and $\Delta$.}
\label{tab:ci-m2}
\begin{tabular}{ c c |c c c c c c | c}\hline \hline
  & &$C^{\rm eff}_7(m_b)$ & $V(m_b)$&$|\overline{D}|^2$&$A(m_b)$&$\Delta$&$R$& ${\cal BR}$ \\
  \hline \hline
 SM&             & $-0.3055$& $-0.024 - 0.015 i$& $0.1089$&$0.0033$&$-0.0012$&$0.1110$&$3.52$ \\ \hline \hline
          &$200$ & $-0.4034$& $-0.004 - 0.020 i$& $0.1661$&$0.0042$&$-0.0037$&$0.1666$&$5.28$\\
          &$250$ & $-0.3857$& $-0.007 - 0.019 i$& $0.1547$&$0.0040$&$-0.0032$&$0.1555$&$4.93$\\
          &$300$ & $-0.3723$& $-0.010 - 0.019 i$& $0.1464$&$0.0039$&$-0.0028$&$0.1475$&$4.68$\\
\cline{2-9}
          &$400$ & $-0.3540$& $-0.014 - 0.018 i$& $0.1354$&$0.0038$&$-0.0024$&$0.1368$&$4.34$\\
Model II  &$600$ & $-0.3345$& $-0.018 - 0.017 i$& $0.1242$&$0.0036$&$-0.0019$&$0.1259$&$3.99$\\
          &$800$ & $-0.3247$& $-0.020 - 0.016 i$& $0.1188$&$0.0035$&$-0.0017$&$0.1207$&$3.83$\\
\cline{2-9}
         &$1000$ & $-0.3192$& $-0.021 - 0.016 i$& $0.1158$&$0.0035$&$-0.0015$&$0.1178$&$3.74$\\
         &$2000$ & $-0.3098$& $-0.023 - 0.015 i$& $0.1109$&$0.0034$&$-0.0013$&$0.1131$&$3.59$\\
         &$3000$ & $-0.3075$& $-0.024 - 0.015 i$& $0.1098$&$0.0034$&$-0.0012$&$0.1120$&$3.55$\\
\hline\hline
\end{tabular}
\end{center}
\end{table}

\begin{figure}[tbh] 
\vspace{2.5cm}
\centerline{\mbox{\epsfxsize=14cm\epsffile{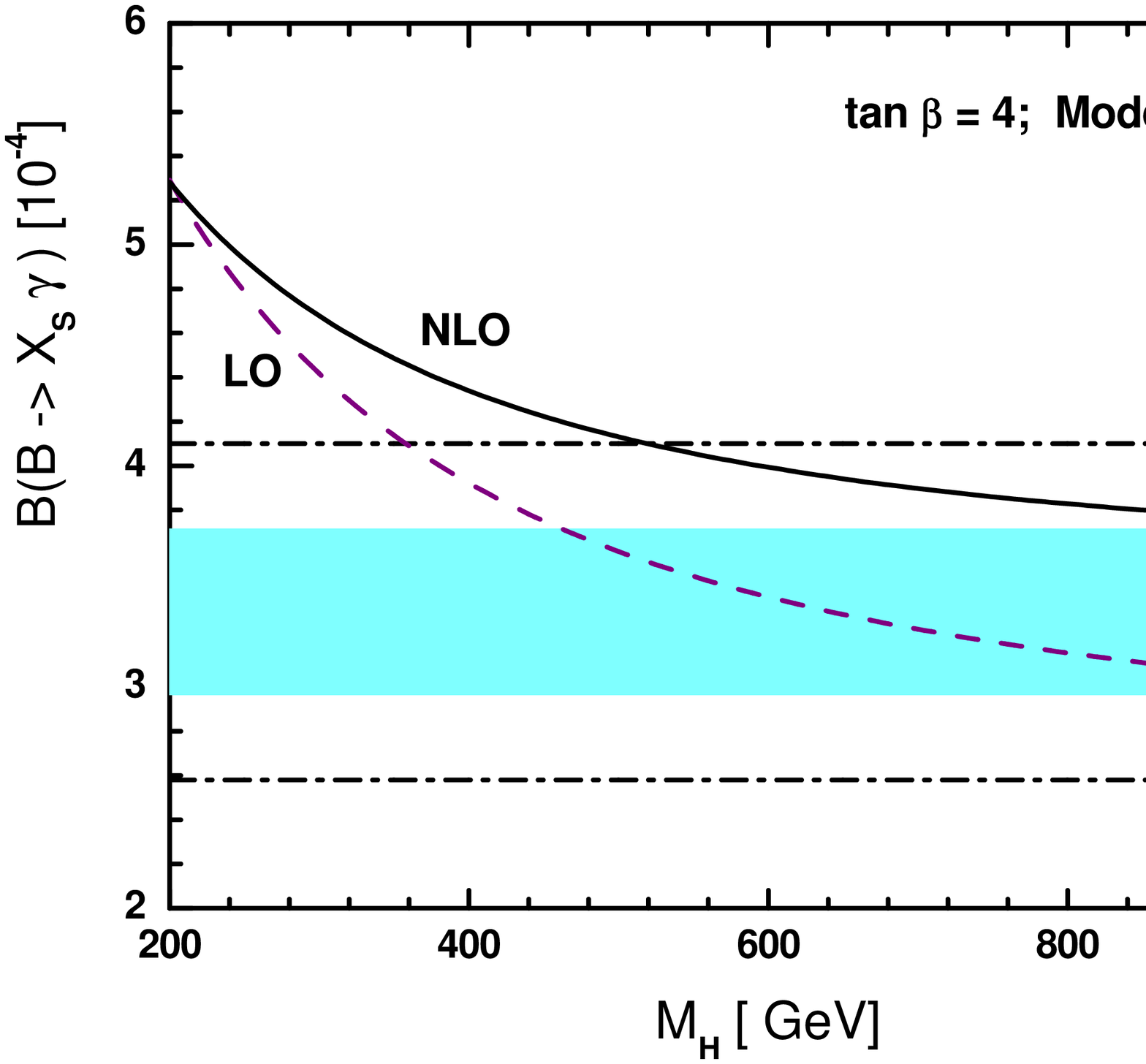}}}
\vspace{-4.5cm}
\caption{ Plots of the branching ratio $\brbxsga$ vs the mass $\mh$ in model II
for $\tan{\beta}=4$. For details see text.}
\label{fig:fig15}
\end{figure}

Fig.~\ref{fig:fig15} shows the $\mh$ dependence of the branching ratio $\brbxsga$ in model II
at LO and NLO level, assuming $\tan{\beta}=4$. The shaded band and the band between two horizontal
dot-dashed lines shows the measured branching ratio $\brbxsga$ within $1\sigma$
and $2\sigma$ errors, respectively.
The dashed and solid curve shows the LO and NLO model II prediction. By comparing the theoretical
predictions and the measured branching ratios at $2\sigma$ level, it is easy to read off the
lower limit on $\mh$ directly from Fig.~\ref{fig:fig15}:
\beq
\mh \geq 357 GeV
\eeq
at LO level,  and
\beq
\mh \geq 520 GeV
\eeq
at NLO level, if the uncertainties of the input parameters are not taken into account.
If we consider the combined uncertainties as given in Eqs.(\ref{eq:brm2lo},\ref{eq:brm2nlo})
the lower limit will be changed to
\beq
\mh \geq 298\ \  {\rm or} \ \  350 \ \  {\rm GeV}
\eeq
at the LO and NLO level, respectively.

\begin{figure}[tbh]
\vspace{2.5cm}
\centerline{\mbox{\epsfxsize=14cm\epsffile{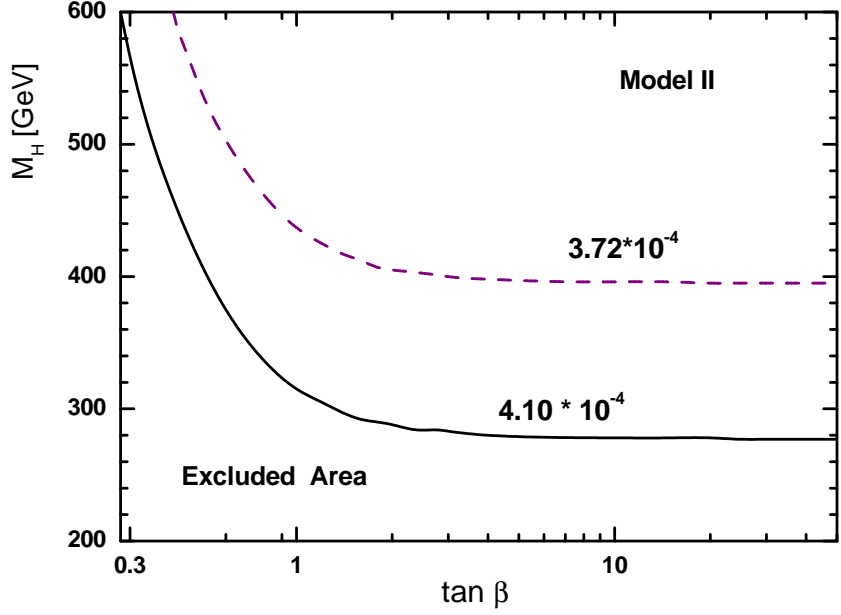}}}
\vspace{-4.5cm}
\caption{ Contour plot in $\tan{\beta}-\mh$ plane obtained by using the NLO model II
predictions and the measured decay rate at $1\sigma$ (dashed curve) and $2\sigma$
level (solid curve). The excluded region is below the corresponding curves.}
\label{fig:fig16}
\end{figure}

Fig.~\ref{fig:fig16} is the contour plot in $\tan{\beta}-\mh$ plane obtained by
finding the minimum of the NLO model II prediction of the branching ratio
$\brbxsga$, when varying the input parameters within their errors and the lower
scale $\mub$ in the range of $2.4 \leq \mub \leq 9.6$ GeV, but fixing
$\muw=\mw$. The excluded region is below the corresponding curves when the measured
branching ratio at $1\sigma$ (dashed curve) and $2\sigma$ level (solid curve)
is employed. For $\brbxsga \leq 4.1 \times 10^{-4}$ and $\tan{\beta} = 1, 10, 20$,
the lower limit on $\mh$ is $315$, $278$, and $277$ GeV, respectively.
Because of the flatness of the curves shown in
Fig.~\ref{fig:fig16} towards the higher end of $\mh$, the lower limits on $\mh$
are very sensitive to the ways to deal with theoretical errors or the details of
the calculations. The lower limit on $\mh$ as illustrated in
Fig.\ref{fig:fig16} is consistent with that given in Ref.\cite{bg98}.

The major differences between model II and the model III studied here and the causes inducing
such differences are the following
\begin{itemize}
\item
In model II, the region of $\tan{\beta} \leq 1$ is strongly disfavored by the measured $\dmd$
and other experiments. In the region $\tan{\beta} \geq 4$, we have $|Y|^2 = 1/\tan^2{\beta}
\ll 1 $ and $XY^* \equiv 1$, the possible new physics contributions
to the Wilson coefficients $C_{7,8}$ are small in magnitude and
have the same sign with their SM counterparts, and therefore
strongly constrained  by the excellent agreement between the data and the SM prediction for
$B \to X_s \gamma$ decay.

\item
In the model III, the new physics contributions proportional to $|\ltt|^2$ and $|\ltt||\lbb|$ have
the opposite sign for $\theta=0^\circ$ and will cancel each other. Even if $|\ltt|$ should
be smaller than $1$ due to the strong constraint from the measured mass
difference $\dmd$ as discussed in last section,
the inclusion of new physics contribution still can change the sign
of the Wilson coefficient $C^{\rm eff}_{7}(m_b)$ from negative in the SM to positive
in model III. On the experimental side, current date still can not exclude the possibility
of a positive Wilson coefficient $C^{\rm eff}_7(m_b)$.

\item
Other NLO terms in the curly bracket of Eq.(\ref{eq:br-nlo}) also receive new physics corrections,
and therefore the cancellation between individual terms  in model III
is very different from the pattern in model II.

\item
The inclusion of NLO contributions will decrease the lower limit on  $\mh$ in model III.
In model II, in contrast, the lower limit will goes up by including the NLO corrections.

\item
In the model III studied here, a light charged Higgs boson with a mass around or even less than
200 GeV is still allowed at NLO level. In model II, however, such light charged
Higgs boson seems impossible.

\end{itemize}

\subsection{Effects of a non-zero phase $\theta$ }

For the mass splitting $\dmd$, the new physics contributions in model III depend on $|\ltt|^2$
and $|\ltt|^4$, and therefore are independent of the phase of coupling $\ltt$.

For the $\bxsga$ decay, however, the dominant new physics contributions ( i.e. the third terms in
Wilson coefficients $C^{0,{\rm eff}}_{7,8}(\mw)$ and $C^{1,{\rm eff}}_{7,8}(\mw)$
in Eqs.(\ref{eq:c70mw},\ref{eq:c80mw}) and (\ref{eq:c71effmw},\ref{eq:c81effmw}))
depend on the relative phase $\theta$ between $\ltt$ and $\lbb$, as can be seen
from Eq.(\ref{eq:xy-m30}). In previous calculations, we always assume $\theta=0^\circ$. Here we
will consider the effect of a non-zero phase $\theta$.

\begin{figure}[tb]
\vspace{3cm}
\centerline{\mbox{\epsfxsize=14cm\epsffile{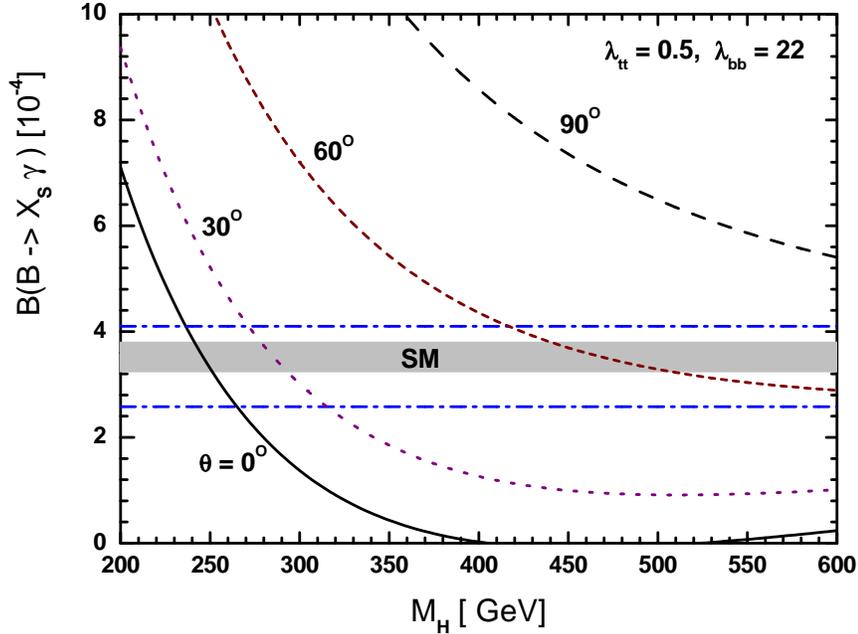}}}
\vspace{-4.5cm}
\caption{ Plots of the branching ratio $\brbxsga$ vs the mass $\mh$ in model III
for case C and for $\theta = 0^\circ$ (solid curve), $30^\circ$ (dots curve),
$60^\circ$ (short-dashed curve) and $90^\circ$ (dashed curve), respectively.
The band between two horizontal dot-dashed lines shows the measured branching ratio within
$2\sigma$ errors. }
\label{fig:fig17}
\end{figure}

Fig.\ref{fig:fig17} shows the $\mh$ dependence of the branching ratio
$\brbxsga$ in model III for case C (i.e.  $(|\ltt|,|\ltt|)=(0.5,22)$)
and for $\theta = 0^\circ$ (solid curve), $30^\circ$ (dots curve),
$60^\circ$ (short-dashed curve) and $90^\circ$ (dashed curve), respectively.
The band between two horizontal dot-dashed lines shows the measured branching ratio within
$2\sigma$ errors. The shaded area shows the SM prediction with the error as given in
Eq.(\ref{eq:brsm}).   It is easy to see from Fig.\ref{fig:fig17} that the non-zero
phase will strengthen the constraint on the mass $\mh$ in case C.
For $\theta=30^\circ$, the allowed region is shifted to
\beq
271 \leq \mh \leq 317 \ \  {\rm GeV}.
\eeq
For  $\theta=60^\circ$, the lower limit on $\mh$ goes up to $417$ GeV. For $\theta \geq
75^\circ$, the lower limit on $\mh$ is higher than $600$ GeV.

\begin{figure}[t] 
\vspace{3cm}
\centerline{\mbox{\epsfxsize=14cm\epsffile{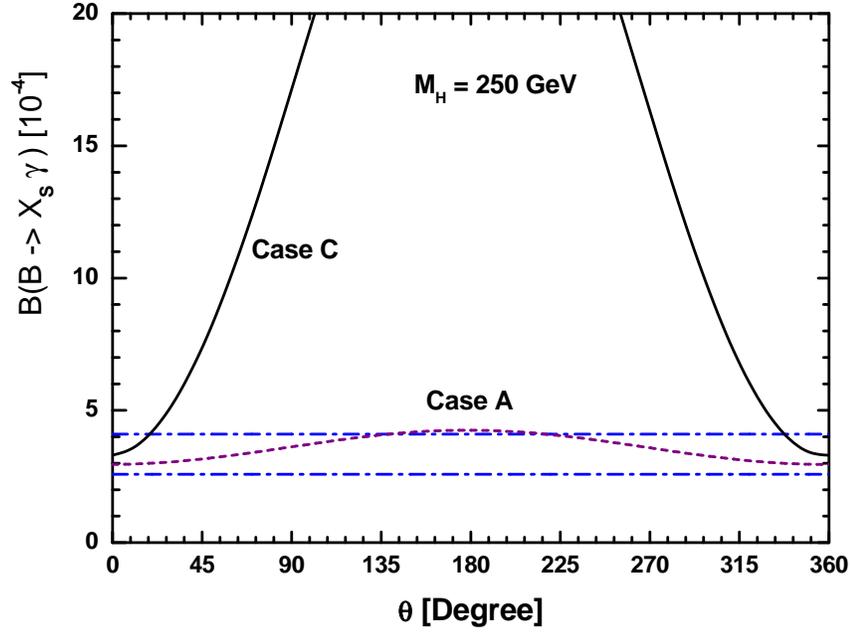}}}
\vspace{-4.5cm}
\caption{ Plots of the branching ratio $\brbxsga$ vs the phase $\theta$ in model III
for the case A (short-dashed curve ) and case C (solid curve), and assuming $\mh=250$
GeV. The band between two horizontal dot-dashed lines shows the measured branching ratio within
$2\sigma$ errors.}
\label{fig:fig18}
\end{figure}

In Fig.\ref{fig:fig18}, we show the $\theta$ dependence of the branching ratio
$\brbxsga$ in model III for the case A (the short-dashed curve) and case C
(the solid curve) discussed in previous subsections, and for given $\mh=250$ GeV.
Obviously, only two narrow regions of $\theta$, $0^\circ \lesssim \theta \lesssim 20^\circ$
and $340^\circ \lesssim \theta \lesssim 360^\circ$, are allowed by the data for
case C. For case A, however, almost the whole ranges of $\theta$ are still
allowed by the data of $\bxsga$ decay.

\section{Summary}

In this paper, we calculated the new physics contributions to the mass splitting $\dmd$
and the branching ratio $\brbxsga$ induced by the charged Higgs loop diagrams
in model III, and found the constraints on the
parameters of model III by comparing the theoretical predictions with the high precision data.
We focus on the effects of the NLO QCD corrections to these two physical observables.

In section \ref{sec-2}, we gave a brief review about the structure of the
general two-Higgs-doublet models and the phenomenological studies about such models.
Following previous works\cite{chao99},
we assume that only the diagonal Yukawa couplings $\ltt$ and $\lbb$  are
non-zero for the model III under consideration in this paper
and  study the effects of NLO new physics contributions.

In section \ref{sec-3}, we calculated the new physics contributions to the mass splitting
$\dmd$ in model III at the NLO level.
The magnitude of Yukawa coupling $\ltt$ is strongly constrained by the precision data
$\dmd = 0.502 \pm 0.007 ps^{-1}$.
As shown in Figs.\ref{fig:fig4} and \ref{fig:fig5}, the upper limit on $\ltt$
is $|\ltt| \leq 1.7$, while the choice of $|\ltt| \approx 0.5$
is favored by the measured $\dmd$ if one requires the charged Higgs boson to be
light, say around $200$ GeV.

In section \ref{sec-4}, we calculated the new physics contributions to the rare decay
$B \to X_s \gamma$ in the model III. The NLO QCD corrections are taken into account here.
The new physics contributions to the Wilson coefficients $C^{0,1}_{7,8}(\mu)$ and
the interference between the new physics parts and their SM counterparts are investigated.
The new physics contributions to $\bxsga$ decay and possible constraints on
$|\lbb|$ and $\mh$ are calculated and analyzed, three typical cases for the choice of Yukawa
couplings $|\ltt|$ and $|\lbb|$ are studied in great detail.
The common features and the differences between the conventional model II and the model
III, as well as the effects of a non-zero relative phase $\theta$ are also considered.
From the numerical results, we found that
\begin{itemize}

\item
In the model III studied here, a light charged Higgs boson with a mass around or even less than
200 GeV is still allowed at NLO level by the measured branching ratio $\brbxsga$ within $2\sigma$
errors, as can be seen in Eq.(\ref{eq:limit}) and Figs.\ref{fig:fig13} and \ref{fig:fig14}.
In model II, however, such light charged Higgs boson seems impossible.

\item
The inclusion of the NLO QCD contributions will decrease the lower limit on  $\mh$ by about
$200$ GeV in model III.
In model II, in contrast, the lower limit on $\mh$ will be increased by about $160$ GeV
because of the inclusion of the NLO corrections.

\item
As illustrated in Fig.\ref{fig:fig17}, the allowed region of $\mh$ in model III
will be shifted toward heavy mass end for a non-zero relative phase
$\theta$.

\end{itemize}

\begin{acknowledgments}

Z.J.Xiao is very grateful to the high energy section of ICTP, Italy, where part
of this work is done, for warm hospitality and financial support.
This work was supported by the National Natural Science Foundation of
China under Grant No.~10075013 and 10275035, and by the Research Foundation
of Nanjing Normal University under Grant No.~214080A916.

\end{acknowledgments}

\begin{appendix}

\section{Input parameters and running couplings} \label{app:input}

In this appendix we list the input parameters used in our calculations. The values of most
parameters are quoted from Ref.\cite{pdg2003,hfag03} directly. The
masses $\mt, m_b, m_c$ are understood to be the pole mass, while $\overline{m}_q(\mu)$
is the running
q-quark mass in the modified minimal subtraction ($\overline{MS}$) scheme at the
renormalization scale $\mu$. To first order in $\alpha_S$, the running mass
$\overline{m}_q(\mu)$ and the pole mass $m_q$ are related through\cite{bg98}
\beq
\overline{m}_q(\mu)= m_q \left [ 1 + \frac{\alpha_S(\mu)}{\pi}
\left ( \ln\frac{m_q^2}{\mu^2} - \frac{4}{3}\right )\right ]
\label{eq:mass}
\eeq

The masses, coupling constants and other input parameters  are \cite{pdg2003,hfag03}
\beq
 M_{B_d} &=& 5.279 GeV, m_b=4.8 \pm 0.2 GeV, \; m_c/m_b= 0.29 \pm 0.02;  \non
\mw &=& 80.42 GeV, \; \mt = 174.3 \pm 5.1 GeV, \non
\alpha_S(\mz) &=& 0.119\pm 0.004, \; \alpha_{em}=1/137.056, \;\non
\lambda_1 &=& -0.50 GeV^2, \; \lambda_2 = 0.12 GeV^2
\label{eq:input}
\eeq

The ratio of CKM elements $|V_{ts}^* {V_{tb}}/V_{cb}|^2$ appeared in the decay
rate $\brbxsga$ can be expressed in terms of the Wolfenstein parameters as follows
\beq
\left | \frac{{V_{ts}^*} {V_{tb}}}{V_{cb}}\right |^2 &=&
1 - \lambda^2 (1-2 \bar{\rho}) + \lambda^4 (\rhob^2 +  \etab^2 -A^2 )
+ {\cal O}(\lambda^6)\non
&=& 0.971 \pm 0.010, \label{eq:ratiov}
\eeq
where we have used $\lambda=0.2196$, $A=0.854$, $\rhob=0.22\pm 0.10$ and
$\etab=0.35 \pm 0.05$ \cite{pdg2003}. The error induced by $\delta \etab = 0.05$
is only $0.0001$ for the ratio.

For the semileptonic branching ratio ${\cal B}_{SL}= \calb (B \to X_c e \overline{\nu}_e)$
we use
\beq
{\cal B}_{SL}=(10.64\pm 0.23)\%
\eeq
as given in Ref.\cite{pdg2003}.

Finally, for the running of $\alpha_s(\mu)$, we use the two-loop formulae as given
in Ref.\cite{buras98}:
\beq
\alpha_s = \frac{\alpha_s(M_Z)}{v(\mu)} \left [ 1-
\frac{\beta_1}{\beta_0} \frac{\alpha_s(M_Z)}{4\pi}
\frac{\ln(v(\mu))}{v(\mu)} \right ],
\eeq
with
\beq
v(\mu) =1 -\beta_0 \frac{\alpha_s(M_Z)}{2\pi}
\ln\left ( \frac{M_Z}{\mu} \right ),
\eeq
where $\beta_0=23/3$ and $\beta_1=116/3$ for b quark decays.

\section{NLO Wilson coefficients at $\mu= \mw$} \label{app-wc}

For the completeness, we list here the NLO functions at the matching scale
$\mu_W$ appeared in Eqs.(\ref{eq:c41effmw},\ref{eq:c71effmw},\ref{eq:c81effmw}).
For more details see Ref.\cite{bg98}.

In the SM, we have
\beq
 E_0(x) & = & \frac{x (x^2+11x-18)}{12 (x-1)^3}
+\frac{x^2 (4 x^2-16x+15)}{6(x-1)^4} \ln x-\frac{2}{3} \ln x -\frac{2}{3}, \\
 W_{7,SM}(x) & = & \frac{-16 x^4 -122 x^3 + 80 x^2 -  8 x}{9 (x-1)^4} \,
  {\rm Li}_2 \left(\!1\!-\!\frac{1}{x} \right) \non
&& +\frac{6 x^4+46 x^3-28 x^2}{3 (x-1)^5} \ln^2 x  \\
                   &   &
+\frac{-102x^5-588 x^4-2262 x^3+3244 x^2-1364 x+208}
    {81(x-1)^5} \ln x                                 \\
                   &   &
+\frac{1646x^4+12205x^3-10740x^2+2509x-436}{486(x-1)^4}, \\
 W_{8,SM}(x) & = & \frac{-4 x^4 +40 x^3 + 41 x^2 + x}{6 (x-1)^4} \,
  {\rm Li}_2 \left(\!1\!-\!\frac{1}{x} \right)
+\frac{ -17 x^3 - 31 x^2}{2 (x-1)^5} \ln^2 x \\
                   &   &
+\frac{-210 x^5+1086 x^4+4893 x^3+2857 x^2-1994 x+280}
     {216 (x-1)^5} \ln x           \\
                   &   &
+\frac{737x^4-14102 x^3-28209x^2+610 x-508}{1296(x-1)^4},  \\
 T_{7,SM} (x)  &=& \frac{x}{3} \, \left[\frac{47x^3-63x^2+9x+7-(18x^3+30x^2-24x)
                       \ln x}{(x-1)^5} \right] ,        \\
 T_{8,SM} (x)  &=&  2x \, \left[\frac{-x^3-9x^2+9x+1+(6x^2+6x)
                       \ln x}{(x-1)^5} \right] \,,
\label{eq:nlo-sm}
\eeq
where $x=m_t^2/\mw^2$.

In the 2HDMs, the NLO functions describing the charged Higgs contributions are
\beq
E_H (y) & = & \frac{y}{36} \, \left[ \frac{7y^3-36y^2+45y-16+(18y-12) \ln y}{(y - 1)^4} \right]
,
\eeq
\beq
 W_{7,YY}(y) & = & \frac{2y}{9}\left[ \frac{8y^3-37y^2+18y}{(y -1)^4}\,
  {\rm Li}_2 \left(\!1\! -\! \frac{1}{y} \right)
+\frac{3y^3+23y^2-14y}{(y -1)^5}\ln^2y \right. \non
&& \left.
+ \frac{21y^4-192y^3-174y^2+251y-50}{9(y-1)^5} \ln y \right.     \non
&   & \left.
 +\frac{-1202y^3+7569y^2-5436y+797}{108(y-1)^4}\right]
 \ - \frac{4}{9} \, E_H    ,
\eeq
\beq
W_{8,YY} (y)& = & \frac{y}{6} \left[ \frac{13y^3-17y^2+30y}{(y-1)^4}\,
 {\rm Li}_2  \left(\!1\! -\! \frac{1}{y} \right)
 -\frac{17y^2+31y}{(y-1)^5}\ln^2y \right. \non
 && \left.
 +\frac{42y^4+318y^3+1353y^2+817y-226}{36(y-1)^5}\ln y \right. \non
 && \left.
+\frac{-4451y^3+7650y^2-18153y+1130}{216(y-1)^4}\right]
 \ - \frac{1}{6} \, E_H  ,
\eeq
\beq
 M_{7,YY} (y)  & = &\frac{y}{27}\left[
  \frac{-14y^4+149y^3-153y^2-13y+31-(18y^3+138y^2-84y) \ln y}{(y-1)^5}
 \right], \\
 M_{8,YY} (y) & = &\frac{y}{36}\left[
 \frac{-7y^4+25y^3-279y^2+223y+38+(102y^2+186y) \ln y}{(y-1)^5}
            \right], \\
T_{7,YY} (y) &=& \frac{1}{3} \, T_{7,SM}(x \to y), \\
T_{8,YY} (y)  &=& \frac{1}{3} \, T_{8,SM}(x \to y) \, ,
\label{eq:nlo-yy}
\eeq
and
\beq
 W_{7,XY}(y) & = & \frac{4 y}{3}\left[  \frac{8y^2-28y+12}{3(y-1)^3} \,
 {\rm Li}_2 \left(\!1\! -\! \frac{1}{y} \right)+
\frac{3y^2+14y-8}{3(y-1)^4}\ln^2y \right. \non
& & \left. + \frac{4y^3-24y^2+2y+6}{3(y-1)^4}\ln y
+\frac{-2y^2+13y-7}{(y-1)^3}\right], \\
 W_{8,XY}(y) & = & \frac{y}{3} \left[  \frac{17y^2-25y+36}{2(y-1)^3}\,
 {\rm Li}_2 \left(\!1\! -\! \frac{1}{y} \right) -
  \frac{17y+19}{(y-1)^4}\ln^2y \right. \non
   &  & \left. +\frac{14y^3-12y^2+187y+3}{4(y-1)^4}\ln y
-\frac{3(29y^2-44y+143)}{8(y-1)^3}\right] ,  \\
 M_{7,XY}(y)& = & \frac{2 y}{9} \, \left[  \frac{-8y^3+55y^2-68y+21-(6y^2+28y-16) \ln y}{(y-1)^4}
 \right], \\
 M_{8,XY}(y) & = & \frac{y}{6} \, \left[ \frac{-7y^3+23y^2-97y+81 +(34y+38) \ln y}{(y-1)^4}
                    \right],   \\
T_{7,XY} (y)&=& \frac{2y}{3} \, \left[ \frac{13y^2-20y+7-(6y^2+4y-4)
                       \ln y}{(y-1)^4} \right],  \\
T_{8,XY}   &=& 2y \, \left[ \frac{-y^2-4y+5+(4y+2) \ln y}{(y-1)^4} \right] \,.
\label{eq:nlo-xy}
\eeq
where $y=\mt^2/\mh^2$.
\end{appendix}

\newpage

\newpage

\end{document}